\documentclass[12pt]{article}

\pdfoutput=1

\usepackage{amsmath,amsfonts,amssymb,amsthm}
\usepackage{setspace,graphicx}
\usepackage{nicematrix}
\usepackage{natbib}
\usepackage{bm}
\usepackage{ifthen}
\usepackage{verbatim}
\usepackage{color}
\usepackage{hyperref}
\usepackage{dsfont}
\usepackage{tikz}
\usepackage{setspace,graphicx}
\usepackage{natbib}
\usepackage{bm}
\usepackage{multirow}
\usepackage{ifthen}
\usepackage{verbatim}
\usepackage{xcolor}
\usetikzlibrary{arrows.meta}
\usepackage{comment}

\theoremstyle{plain}
\theoremstyle{definition}
  \newtheorem{theorem}{Theorem}

  \newtheorem{claim}{Claim}
  
  \newtheorem{corollary}{Corollary}
  \newtheorem{definition}{Definition}
  
  \newtheorem{example}{Example}
  \newtheorem{lemma}{Lemma}
  
  \newtheorem{proposition}{Proposition}
  \newtheorem{remark}{Remark}
  \theoremstyle{remark}

\DeclareMathOperator{\argmax}{argmax}

\DeclareMathOperator{\support}{supp}
\DeclareMathOperator{\convexhull}{conv}

\title{Contracting with Heterogeneous Researchers}
\author{Han Wang\thanks{\href{mailto:wang.12716@osu.edu}{wang.12716@osu.edu}; Department of Economics, The Ohio State University\\
I want to thank Yaron Azrieli, Paul J. Healy, Yonggyun Kim, Dan Levin, Mich\`ele M\"uller-Itten, James Peck, John Rehbeck, Huanxing Yang, Renkun Yang, Nathan Yoder, Kun Zhang and participants at the OSU Theory/Experimental reading group, the 2022 Stony Brook International Conference on Game Theory, the Fall 2022 Midwest Economic Theory Conference and the 2023 Annual Meeting of Midwest Economics Association for their valuable comments. All remaining errors are mine.}}
\date{}

\begin{document}

\maketitle

\begin{abstract}
We study the design of contracts that incentivize a researcher to conduct a costly experiment, extending the work of \cite{Y2022} from binary states to a general state space. The cost is private information of the researcher. When the experiment is observable, we find the optimal contract and show that higher types choose more costly experiments, but not necessarily more Blackwell informative ones. When only the experiment result is observable, the principal can still achieve the same optimal outcome if and only if a certain monotonicity condition with respect to types holds. Our analysis demonstrates that the general case is qualitatively different than the binary one, but that the contracting problem remains tractable.

 \bigskip



\pagebreak
\end{abstract}

\newpage

\section{Introduction}

 Acquiring information through contracting is a prevalent practice when facing significant decisions. Governments may engage experts to evaluate vaccine efficacy, while business owners may consult with economists to design investment plans. The principal relies on the agent to gather information in the hopes of making more informed decisions. For instance, in vaccine testing, the agent has a significant degree of flexibility in selecting experimental designs but must comply with legal requirements, such as pre-registering trials and publicly disclosing their results. Despite having hard information from the researcher's experiment, two major concerns remain regarding the agent's incentives: Firstly, the agent may have more information about the cost of acquiring information than the principal. Secondly, the principal may lack the ability to observe or validate the research method based solely on the observed result.

We study a principal-agent model of information acquisition, motivated by the example of vaccine testing. The principal seeks to hire a researcher to gather information about the state of the world. Although the researcher has no intrinsic preferences over the decision being made, the principal can incentivize the researcher to design an experiment by offering payments before the true state is revealed. We assume that the researcher can choose any experiment at a private cost. The timing of the model is as follows: the principal offers a contract, and the researcher decides whether to accept or reject the offer. If the offer is accepted, the researcher commits to a stochastic information structure, which will generate a public signal. Finally, the principal chooses an action based on the information revealed by the signal.

This paper builds on the work of \cite{Y2022}, who analyzed the optimal contract with a stylized structure of two states and binary signals. We broaden this work by allowing for more states, which is necessary for many real-world decision problems. Our analysis shows that including more states can significantly affect the results, but the contracting problem remains tractable. Our findings are particularly relevant in many applications, as numerous decision problems involve more than two states. For example, in vaccine testing, the goal is often to determine the possible side effects, which can involve multiple states. Similarly, companies purchasing consumer data from a platform may need to segment the population based on demographics, location, and browsing history. Each segment can represent a different state in such cases, which can significantly impact the optimal contract. 

This paper assumes that research results are observable and contractible. We distinguish between two environments based on the observability of research methods: the \emph{methods-based contracting} environment, where research methods are observable to the principal, and the \emph{results-based contracting} environment, where research methods are not observable. This distinction helps address the hidden-action problem and facilitates effective comparison. We will consider the optimal contract within the most general class under methods-based contracting and explore whether the induced outcome can be achieved using the more restricted class of results-based contracts.

Under methods-based contracting, our goal is to determine the specific experiment choice function that the principal wants to implement. When there are only two states, \cite{Y2022} establishes a condition for the optimal experiment choice function, called ``Blackwell monotonicity''. This means that higher types choose more informative experiments in the Blackwell order, thus simplifying the principal's contracting problem. However, in scenarios involving more than two states, we present an example that shows the optimal experiment choice function may not exhibit Blackwell monotonicity. The intuition is that lower costs make it better to choose more extreme posteriors. With two states, posteriors move in one dimension, but with more than two states, posteriors can ``rotate'' towards the simplex's corner in a way that potentially violates Blackwell monotonicity. Nevertheless, we establish a weaker monotonicity condition, asserting that higher types opt for more costly experiments, which can still streamline the principal's contracting problem.

In the context of results-based contracting, we first explore results-based contracts that incorporate type revelation, departing from \cite{Y2022}'s emphasis on payments that depend solely on research results. The principal gains greater flexibility in designing incentives by permitting payments to be conditional on the reported type. This provides an explanation for budget proposals in grant applications. Our finding connects \citeauthor{Y2022}'s work with \cite{RS2017} and demonstrates that we can construct an outcome-equivalent direct contract with results-based payments for any incentive-compatible methods-based contract. 

We then study results-based contracts without type revelation. Analyzing this specific class of contracts is relevant for many applications as real-world contracts may lack a screening mechanism. When there are only two states, \cite{Y2022} shows that any binary, Blackwell-monotone experiment choice function could be implemented through results-based contracts and hence that the outcome of the optimal methods-based contract can be achieved. Blackwell monotonicity is also necessary for implementation. 

To accommodate more than two states, we define a novel monotonicity concept, which depends on the cost function. We show that this concept characterizes the existence of a results-based contract outcome-equivalent to the optimal methods-based contract. As a result, we can identify when the principal can focus on results-based contracts without any loss, which is helpful in many applications that involve multiple states. An intriguing example arises when the principal wants to implement a ``symmetric'' experiment, that is, one where the cost of generating each piece of evidence is the same. Moreover, we highlight the connections between different notions of monotonicity and illustrate the construction of such contract for scenarios involving two types. Our construction remains applicable even when the optimal experiment choice function does not exhibit Blackwell monotonicity. Overall, our work extends the literature on learning incentives and offers insights for devising effective and flexible contracts in various settings.

\textbf{Related Literature.} --- This paper belongs to the literature on contracting for acquiring information, which has been extensively studied in various settings. Early work includes \cite{O1989}, where the agent incurs a cost for each observation he draws while the principal is unaware of this cost and the number of observations made. \cite{Y2022} builds upon this setting by modeling the agent as an information designer and examining the principal's contracting problem with adverse selection and moral hazard. In a related setting with only adverse selection, \cite{M2021} examines cases where the agent has private information about the set of feasible experiments, and the principal cannot make payments. \cite{RS2017} and \cite{WZ2022} study a model with only moral hazard, where the principal cannot observe the agent's choice of information structure. Both of them address the impacts of risk aversion and limited liability.

Our paper shares the assumption with \cite{RS2017} and \cite{Y2022} that research results are observable and contractible. These two papers provide valuable insights into how to incentivize the researcher in this setting. Nevertheless, incentivizing information acquisition can be challenging when research results are not verifiable. Several papers have proposed solutions to this problem. One approach is to provide incentives based on the ex-post evaluation of the researcher's advice, such as the realized state or the ex-post payoff of the principal. Examples of such work include \cite{Z2011}, \cite{HT2018}, \cite{C2019}, \cite{CR2021}, and \cite{WZ2022}. Another approach is to consider peer monitoring with multiple experts. For instance, \cite{A2021,A2022} examine this case by letting payments contingent on the entire vector of reports.

In our model, it is the decision maker who designs contracts for procuring information. Recent research has also explored situations where a data broker offers contracts in order to sell information, as surveyed by \cite{BB2019}. For example, \cite{BBS2018} examines a model in which a data broker sells experiments to a decision-maker who has private information about the prior belief. \cite{Y2022} investigates a setting where a data broker sells market segmentations to a firm with private cost. \cite{L2021} explores the data broker's optimal selling mechanism when the decision maker has the option to conduct an experiment at an additional cost. 

Methodologically, this paper is related to mechanism design and information design. See, for instance, \cite{M1981,M1982}, \cite{MR1984}, \cite{KG2011}, \cite{BM2019} and, \cite{K2019}. Our model differs from Bayesian persuasion in assuming the agent has no stake in the principal's decision problem. As a result, the agent designs the information to maximize the payment net cost rather than persuades the decision-maker to choose a preferred action. Furthermore, specifying experiments for different types of the researcher can be viewed as a comparative statics question within the literature of rational inattention, as examined by \cite{MM2015} and \cite{CDL2017,CDL2019}.

\section{Model Setting}

\subsection{Notations}
We write $\mathbb{R}_{+}$ for the set of non-negative real numbers and $\mathbb{R}_{++}$ for the set of positive real numbers. The notation $conv(S)$ for a set $S\subseteq \mathbb{R}^{k}$ denotes the convex hull of $S$. 

Let the space of probability distributions over a finite set $S$ be $\Delta(S)=\{p\in\mathbb{R}_{+}^{|S|}: \sum_{s\in S}p(s)=1\}$, where $|S|$ is the cardinality of $S$. For a distribution $p \in \Delta(S)$, we write $supp(p)=\{s\in S: p(s)>0\}$. A full support distribution $p$ has $p(s)>0$ for all $s\in S$.

\subsection{Setup}
We extend the principal-agent model of information acquisition presented in \cite{Y2022} to accommodate more than two states. 

The principal (she) is faced with a Bayesian decision problem and can hire a researcher (he) to learn about the uncertain state of the world. Suppose that the state of the world $\omega$ can take values in a finite set $\Omega$. The principal's Bayesian decision problem is a triplet $D=(A, u, p_{0})$, including a finite set of actions $A$, a utility function $u: \Omega \times A\to \mathbb{R}$ and a full support prior belief over the states $p_{0}\in \Delta(\Omega)$. We let $v(p):=\max_{a\in A}\left[\sum_{\omega\in \Omega} p(\omega)\cdot u(\omega,a)\right]$ be the maximal achievable expected utility at belief $p\in\Delta(\Omega)$. 

The researcher has the same prior belief $p_{0}$. Once hired, he can choose any experiment to learn about the state. After observing the signal realization, everyone updates the belief according to Bayes rule. Under the common prior assumption, it is convenient to represent signal structures as distributions over posteriors that they induce. Given $p_{0}$, for any distribution over posteriors $\tau$, there exists a signal structure inducing it if and only if $\tau$ satisfies $\mathbb{E}_{p\sim\tau}\left[p\right]=p_{0}$ \citep{AM1995,KG2011}. The requirement that the expectation over posteriors equals the prior is known as \emph{Bayes plausibility}. We denote by $X(p_{0}) \subseteq \Delta\left(\Delta\left(\Omega\right)\right)$ the set of distributions over posteriors that have finite support and satisfy Bayes plausibility. We will write $X$ instead of $X(p_{0})$ to simplify the notation. Let an experiment be $\tau\in X$. We also call an experiment a \emph{research method} and refer to the realized posterior as its \emph{result}. 

The researcher is better informed about the cost of acquiring information. He has a private type $\theta\in \Theta$, and the cost of experiment $\tau$ is  $\theta \cdot C(\tau)$. Consider $ \Theta:=\{\theta_{1},\theta_{2},\ldots,\theta_{N}\}\subseteq \mathbb{R}_{++}$, where $\theta_{1}>\theta_{2}>\ldots>\theta_{N}$. A larger subscript is associated with a more efficient type that has a lower cost of running experiments. Type $\theta$ is drawn from a commonly known distribution with full support. Let $F$ denote the cumulative distribution function and $f$ denote the corresponding probability mass function. We assume the cost $C(\tau)$ is posterior-separable \citep{CDL2017}, i.e., there is a continuous and strictly convex function $c: \Delta(\Omega)\to \mathbb{R}_{+}$ with $c(p_{0})=0$ such that $C(\tau)=\mathbb{E}_{p\sim \tau}\left[c(p)\right]$. In this paper, we will take $c$ as a primitive of the model, while conveniently representing $\mathbb{E}_{p\sim \tau}\left[c(p)\right]$ as $C(\tau)$.  
 
This paper considers a setting with quasi-linear expected-utility preferences. An outcome is a pair $(\tau, \phi)\in X \times \mathbb{R}$ that specifies an experiment and a transfer. The principal's payoff is $\mathbb{E}_{p\sim\tau}\left[v\left(p\right)\right]-\phi$, where $\mathbb{E}_{p\sim\tau}\left[v\left(p\right)\right]$ is the expected utility from decision problem $D$ under experiment $\tau$. The researcher runs an experiment at some cost. A type-$\theta$ researcher's payoff is given by $\phi-\theta \cdot C(\tau)$. We assume that the researcher has no stake in the decision problem $D$; he only cares about the payment from the principal and the experiment's cost. 

The principal maximizes her expected payoff by making a take-it-or-leave-it offer to the researcher. We take the principal's perspective and study the design of optimal contracts. In line with \cite{Y2022}, we assume the experiment's results are observable and contractible. We consider two distinct environments: \emph{methods-based contracting}, where the principal is capable of observing the choice of experiments, and \emph{results-based contracting}, where such observation is not feasible.

\section{Methods-based Contracting}

In this section, we study a benchmark model assuming that research methods are observable and contractible. By the revelation principle, we can restrict attention to the class of direct revelation contracts, namely ``methods-based contracts''. We focus on how to design incentives for the researcher to report his type truthfully. We generalize the analysis in \cite{Y2022} to accommodate more complicated $D$, particularly when the world has more than two states.

 We define a methods-based contract as a pair of functions $(\mathcal{X},T)$, where $\mathcal{X}: \Theta\to X$ is an experiment choice function and $T: \Theta\to \mathbb{R}$ is a payment function. If the researcher reports truthfully, the principal's payoff is $$U_{P}(\mathcal{X},T)=\mathbb{E}_{\theta \sim F}\left[\mathbb{E}_{p\sim \mathcal{X}\left(\theta\right)}\left[v(p)\right]-T(\theta)\right].$$ 
 
 The payoff of a type-$\theta$ researcher reporting $\theta^{\prime}$ is $$U_{\theta}(\theta^{\prime}|\mathcal{X},T)=T(\theta^{\prime})-\theta \cdot C(\mathcal{X}\left(\theta^{\prime}\right)).$$ 

We say that a methods-based contract $(\mathcal{X},T)$ is \emph{incentive-compatible (IC)} if every type of researcher prefers truthful reporting, i.e., $\forall \theta\in\Theta$, $\theta\in \argmax_{\theta^{\prime}\in \Theta} U_{\theta}(\theta^{\prime}|\mathcal{X},T)$. We say that $(\mathcal{X},T)$ is \emph{individually rational (IR)} if every type of researcher prefers to accept the contract upon truthful reporting, i.e., $\forall \theta\in\Theta$,  $U_{\theta}(\theta|\mathcal{X},T)\geq 0.$ 

The principal searches over the set of IC and IR methods-based contracts to maximize her ex-ante payoff. The \emph{methods-based contracting} problem is defined as follows.
\begin{align}
    &\max_{\mathcal{X},T} \quad U_{P}(\mathcal{X},T)\notag\\ 
    \text{s.t. } & (\mathcal{X},T) \text{ is IC and IR}\tag{Methods}\label{mbc} 
\end{align}

Before the analysis, let us introduce two definitions of monotone choice functions. 

\begin{definition} Consider an experiment choice function $\mathcal{X}:\Theta \to X$.\hfill 
\begin{enumerate}
    \item  $\mathcal{X}$ is \emph{$c$-monotone}, if $C(\mathcal{X}(\theta_{j}))\geq C(\mathcal{X}(\theta_{i}))$, $\forall i,j \text{ with } j>i$. 
    
\item  $\mathcal{X}$ is \emph{Blackwell-monotone}, if it is \emph{$c$-monotone} for every convex and continuous function $c: \Delta(\Omega)\to \mathbb{R}$. \citep{B1951,B1953} 
\end{enumerate}

\end{definition}

These two monotonicity concepts rely on different orderings of experiments. It is known that the definition of Blackwell monotonicity is equivalent to saying that $\mathcal{X}(\theta_{j})$ is more informative than $\mathcal{X}(\theta_{i})$ for all $i,j$ with $j>i$. Note that $c$-monotonicity is a weakening of Blackwell monotonicity. While $c$-monotonicity depends on the function $c$, Blackwell monotonicity does not.  

As shown in \cite{Y2022}, any incentive-compatible contract leads to $c$-monotonicity and the principal's methods-based contracting problem has much in common with the classic adverse selection problem \citep{MR1984}. Define a \emph{virtual type} function $g(\theta)$ by $g\left(\theta_{N}\right)= \theta_{N}$ and $g\left(\theta_{k}\right) =\theta_{k}+\frac{F\left(\theta_{k+1}\right)}{f\left(\theta_{k}\right)}\left(\theta_{k}-\theta_{k+1}\right)$ for $k\leq N-1$. We can rewrite the principal's methods-based contracting problem as follows. 

\begin{lemma}[Yoder, 2022] \hfill \label{lemma:1}

$(\mathcal{X}^{*}, T^{*})$ solves the principal's problem (\ref{mbc}) if and only if the following conditions hold: 
\begin{align}
    \mathcal{X}^{*} \in \underset{\mathcal{X}}{\argmax} &\quad \mathbb{E}_{\theta}\left[\mathbb{E}_{p\sim \mathcal{X}\left(\theta\right)}\left[v(p)-g(\theta)c(p)\right]\right]\label{eq:M}\\
    &\text{s.t. } \mathcal{X} \text{ is $c$-monotone}\notag\\
    T^{*}\left(\theta_{1}\right)&= \theta_{1} C\left(\mathcal{X}^{*}\left(\theta_{1}\right)\right) \label{eq:Tstar1}\\
    T^{*}\left(\theta_{k}\right)&= \theta_{k} C\left(\mathcal{X}^{*}\left(\theta_{k}\right)\right)+\sum_{i=1}^{k-1}\left(\theta_{i}-\theta_{i+1}\right) C\left(\mathcal{X}^{*}\left(\theta_{i}\right)\right) \quad \text{for } k\geq 2 \label{eq:Tstar2}
\end{align}

Moreover, if $T^{*}$ is derived from $\mathcal{X}^{*}$ according to (\ref{eq:Tstar1}) and (\ref{eq:Tstar2}), then among all IC and IR methods-based contracts $(\mathcal{X}^{*},T)$,  $T^{*}(\theta)\leq T(\theta)$, $\forall \theta\in \Theta$. 
 
\end{lemma}

\begin{proof}[Proof of Lemma \ref{lemma:1}]
See Lemma 4 in \cite{Y2022}.
\end{proof}
 
 Lemma \ref{lemma:1} provides a tractable way of finding an optimal methods-based contract. We can solve $\mathcal{X}^{*}$ from the constrained problem (\ref{eq:M}) and derive the corresponding $T^{*}$ by applying (\ref{eq:Tstar1}) and (\ref{eq:Tstar2}) to $\mathcal{X}^{*}$. Notably, $T^{*}$ is the cheapest way to implement $\mathcal{X}^{*}$. 
 
 The constrained problem (\ref{eq:M}) features an additively separable objective in $\theta$, while the $c$-monotonicity constraint imposes requirements on experiments across different types. Our approach begins with considering a relaxed program that omits the $c$-monotonicity constraint, making it easier to solve. Subsequently, we will verify that the solution derived from the relaxed program satisfies the $c$-monotonicity constraint. 
 
 We define the relaxed problem as follows. It can be written as a set of type-specific problems, i.e., choosing an experiment for each $\theta$.
  \begin{align}
  &\mathcal{X}^{*} \in \underset{\mathcal{X}}{\argmax} \quad \mathbb{E}_{\theta}\left[\mathbb{E}_{p\sim \mathcal{X}\left(\theta\right)}\left[v(p)-g(\theta)c(p)\right]\right]\notag\\
    \iff &\mathcal{X}^{*}(\theta)\in \underset{\tau\in X}{\argmax} \quad \mathbb{E}_{p\sim \tau}\left[v(p)-g(\theta)c(p)\right] \quad \forall \theta \in \Theta\label{eq:type-specific}
\end{align}

 The following proposition shows that we can often safely ignore the $c$-monotonicity constraint. Any solution to the relaxed program must satisfy $c$-monotonicity, then it also solves the constrained problem (\ref{eq:M}).

\begin{proposition} \label{proposition:1} 

If the virtual type $g(\theta)$ is strictly increasing in $\theta$, then any experiment choice function that solves the relaxed program must be $c$-monotone.
\end{proposition}

\begin{proof}[Proof of Proposition \ref{proposition:1}]

Consider $\theta,\theta^{\prime}\in\Theta$ such that $\theta< \theta^{\prime}$. If virtual type $g$ is strictly increasing, then $g(\theta)<g(\theta^{\prime})$. Recall that we interpret $\theta$ as the more efficient type in $\{\theta,\theta^{\prime}\}$. To show that $\mathcal{X}^{*}$ is $c$-monotone, we need $ C(\mathcal{X}^{*}(\theta))\geq C(\mathcal{X}^{*}(\theta^{\prime}))$. Suppose that $ C(\mathcal{X}^{*}(\theta))<C(\mathcal{X}^{*}(\theta^{\prime}))$.

By the optimality of $\mathcal{X}^{*}$, experiment $\mathcal{X}^{*}\left(\theta^{\prime}\right)$ solves the type-specific problem for $\theta^{\prime}$. This implies that  $\mathbb{E}_{p\sim \mathcal{X}^{*}\left(\theta^{\prime}\right)}\left[v(p)-g(\theta^{\prime})c(p)\right]\geq \mathbb{E}_{p\sim \mathcal{X}^{*}\left(\theta\right)}\left[v(p)-g(\theta^{\prime})c(p)\right]$.
    
Rearranging terms and using $g(\theta)<g(\theta^{\prime})$, we get 
\begin{align*}
    \mathbb{E}_{p\sim \mathcal{X}^{*}\left(\theta^{\prime}\right)}\left[v(p)\right]-\mathbb{E}_{p\sim \mathcal{X}^{*}\left(\theta\right)}\left[v(p)\right]
    \geq &  g(\theta^{\prime})\left[C(\mathcal{X}^{*}(\theta^{\prime}))- C(\mathcal{X}^{*}(\theta))\right]\\
    > & g(\theta)\left[C(\mathcal{X}^{*}(\theta^{\prime}))- C(\mathcal{X}^{*}(\theta))\right]
\end{align*}

This can be written as $\mathbb{E}_{p\sim \mathcal{X}^{*}\left(\theta^{\prime}\right)}\left[v(p)-g(\theta)c(p)\right]> \mathbb{E}_{p\sim \mathcal{X}^{*}\left(\theta\right)}\left[v(p)-g(\theta)c(p)\right]$,  meaning that $\mathcal{X}^{*}\left(\theta\right)$ does not solve the type-specific problem at $\theta$. This contradicts the optimality of $\mathcal{X}^{*}$.
\end{proof}
 
Proposition \ref{proposition:1} verifies that the candidate experiment choice function is indeed optimal. It is based on the intuition that people will demand more information, as the marginal cost goes down. The proof of this proposition holds for any finite state space. It is worth noting the connection to \cite{Y2022}: With $|\Omega|= 2$, \cite{Y2022} proves a stronger statement, namely, any choice function that solves the relaxed program must be Blackwell-monotone. However, this statement does not hold in the case of $|\Omega|> 2$.\footnote{For state spaces with $|\Omega|\geq 2$, the impact of the marginal cost parameter on the choice of experiments has been examined by \cite{D2022} and \cite{Y2022}. According to their results, a more efficient type will choose more extreme posteriors. This result is stated formally in Proposition 3 of \cite{D2022} and Proposition 4 of \cite{Y2022}. We will discuss how our results relate to theirs in the Appendix \ref{section:appB}.} Here is an example with three states, where the experiment choice function that solves the relaxed program is not Blackwell-monotone.

\begin{example}\label{example:1}

Suppose that there are three states, $\Omega=\{\omega_{1},\omega_{2},\omega_{3}\}$. Let the action space be $A=\{a_{1},a_{2},a_{3}\}$ and the prior belief be $p_{0}=(\frac{1}{3},\frac{1}{3},\frac{1}{3})$. The utility function $u(a,\omega)$ is described using the following table. 
			
			$$ \begin{array}{|c|c|c|c|}
			\hline
			& \omega_{1} & \omega_{2} & \omega_{3}\\
			\hline
			a_{1}& 5 & 4 & 2 \\
			\hline
			a_{2}& 0 & 5 & 3\\
			\hline
			a_{3}& 5 & 1 & 5\\
			\hline
			\end{array}$$

The researcher has two possible types, $\Theta=\{\theta_{1},\theta_{2}\}$, where $\theta_{1}=\frac{9}{4}$, $\theta_{2}=2$, and $f(\theta_{1})=f(\theta_{2})=\frac{1}{2}$. We can calculate the virtual type $g(\theta)$: $g(\theta_{1})=\frac{5}{2}$ and $g(\theta_{2})=2$. Let the information cost be the reduction in Shannon entropy $c(p)=H(p_{0})-H(p)$, where  
    $H(p)=-\sum_{\omega\in \Omega}p\left(\omega\right)\log(p\left(\omega\right))$. 

We solve the optimal choice function using the characterization in \cite{MM2015} and \cite{CDL2019}. See Figure \ref{figure:1} for a representation in the belief simplex. We defer the numerical results to the Appendix \ref{section:appA}.

\begin{figure}[http]
    \centering
    \includegraphics[width=0.7\textwidth]{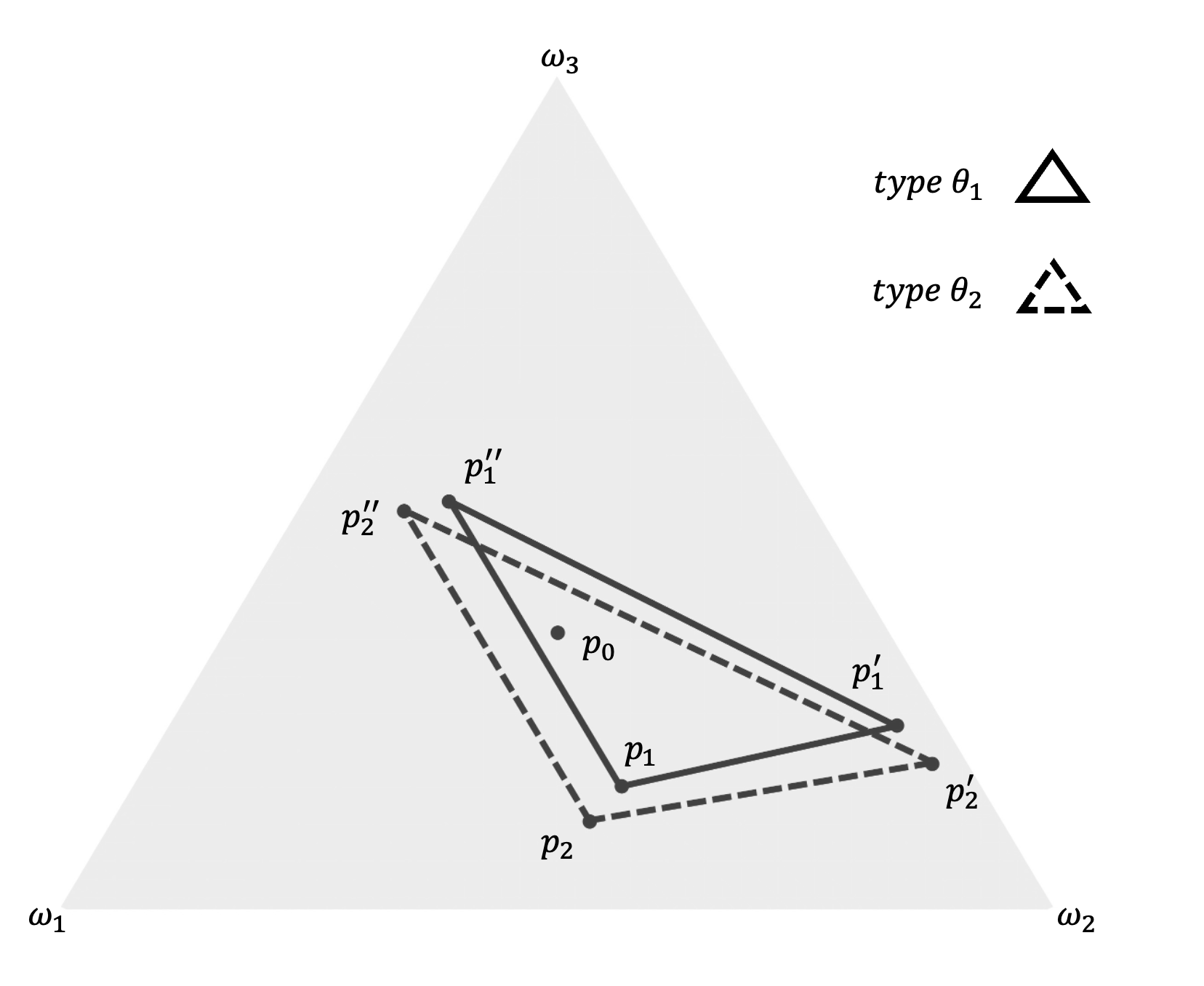}
    \caption{The optimal experiment choice function is not Blackwell-monotone.}
    \label{figure:1}
\end{figure}
    
 Note that for any finite $\Omega$, given two distributions over posteriors  $\tau$, $\tau^{\prime}\in X$, $\tau$ is more informative than $\tau^{\prime}$ in the Blackwell order, only if $\support(\tau^{\prime})\subseteq \convexhull(\support(\tau))$. From Figure \ref{figure:1}, for $\mathcal{X}^{*}$ to be Blackwell-monotone, every vertex of the thick-lined triangle should be inside the dashed-lined triangle. However, we find two vertices of the thick-lined triangle that are not (i.e., $p_{1}^{\prime}$ and $p_{1}^{\prime\prime}$). Therefore, $\mathcal{X}^{*}$ does not satisfy Blackwell monotonicity.

 With a higher virtual type, every posterior will move towards a corner of the simplex. The simplex is a one-dimensional line segment when there are two states, limiting posteriors to move in only one dimension. However, with more than two states, the simplex becomes more complex and enables movement in multiple dimensions. 

\end{example}

\section{Results-based Contracting}

In Section 3, we assumed that the principal could observe and design contracts based on the researcher's research method. However, due to the research method's stochastic nature, only one posterior can be realized from an experiment. One may wonder what happens if the principal has no way to verify the choice made by the researcher. We will assume that research methods are not observable. The realized posterior is observable and contractible, so payments can be made contingent on it.

\subsection{Results-based payments with screening}
We start by analyzing direct contracts in a setting similar to \cite{M1982}, where the agent has both private information and private decisions. In a direct contract, the principal suggests the choice of experiment based on the reported type and pays as a function of both the reported type and the realized posterior. As before, an experiment choice function is a map $\mathcal{X}:\Theta\to X$. A contingent payment function is a map $t:\Theta\times \Delta(\Omega) \to \mathbb{R}$. If the researcher reports his type to be $\theta\in \Theta$ and the realized posterior is $p\in \Delta(\Omega)$; then the researcher will get payment $t(\theta,p)\in \mathbb{R}$. 

We say that $(\mathcal{X},t)$ is incentive-compatible if
\begin{align}\label{newic}
    \mathbb{E}_{p\sim \mathcal{X}(\theta)}[t(\theta,p)]-\theta \mathcal{C}(\mathcal{X}(\theta))\geq\mathbb{E}_{p\sim \tau}[t(\theta^{\prime},p)]-\theta \mathcal{C}(\tau) \quad \forall \theta,\theta^{\prime}\in \Theta, \forall \tau\in X.
\end{align}

Incentive compatibility involves two constraints: obedience and honesty. Obedience means that the researcher follows the suggested experiment choice function, and honesty requires that the researcher reports truthfully. When research methods are not observable to the principal, the researcher may choose outside the set of suggested experiments, i.e., $\{\mathcal{X}(\theta):\theta\in \Theta\}$. The incentive compatibility condition ensures that it should not be a profitable deviation for the researcher to choose any $\tau\in X$.

We will show that given certain regularity conditions, achieving the outcome of any incentive-compatible methods-based contract is possible by using results-based payments with screening. Additionally, We will show that an optimal methods-based contract must satisfy this regularity condition. Therefore, the principal can attain the same expected payoff as under methods-based contracting.

\begin{definition}
    Experiment choice function $\mathcal{X}$ is \emph{non-redundant} if $\support (\mathcal{X}(\theta))\subseteq \Delta(\Omega)$ is affinely independent, $\forall \theta\in\Theta$.
\end{definition}

\begin{theorem}\label{theorem:1}
    Suppose that $\mathcal{X}$ is non-redundant. For any methods-based contract $(\mathcal{X},T)$ that is incentive-compatible, there is a contingent payment rule $t:\Theta\times \Delta(\Omega) \to \mathbb{R}$, such that $(\mathcal{X},t)$ is incentive-compatible and $\mathbb{E}_{p\sim \mathcal{X}(\theta)}[t(\theta,p)]=T(\theta)$.
\end{theorem}

\begin{proof}[Proof of Theorem \ref{theorem:1}]
By affine independence, $|\support (\mathcal{X}(\theta))|\leq|\Omega|$, $\forall \theta \in\Theta$. It follows that for given $\theta$, there exists a hyperplane that contains all points in $\{(p,c(p)):p\in \support (\mathcal{X}(\theta))\}$.\footnote{We can select one arbitrarily if there are multiple hyperplanes that contain all these points. In addition to affine independence, if $|\support (\mathcal{X}(\theta))|=|\Omega|$, then such hyperplane is uniquely pinned down by $\{(p,c(p)):p\in \support (\mathcal{X}(\theta))\}$.} We can identify this hyperplane by an affine function $H_{\theta}:\Delta(\Omega)\to \mathbb{R}$. Set $\underline{T}=\min_{\theta\in\Theta}\left[\min_{p\in\Delta(\Omega)}\left[T(\theta)-\theta H_{\theta}(p)\right]\right]$. 

We can define a contingent payment rule as \begin{align*}
        t(\theta,p)=\begin{cases}
            T(\theta) \quad \text{if } p\in \support(\mathcal{X}(\theta))\\
            \underline{T} \quad \text{if } p\not\in \support(\mathcal{X}(\theta))\\
        \end{cases}, \forall \theta\in \Theta.
    \end{align*}

  Immediately, $\mathbb{E}_{p\sim \mathcal{X}(\theta)}[t(\theta,p)]=T(\theta)$ holds from the definition. 
  
  We need to show that $(\mathcal{X},t)$ is incentive-compatible.
  
  \textbf{Obedience} Upon reporting $\theta$, the researcher's payoff of choosing any experiment $\tau\in X$ is bounded by that of choosing $\mathcal{X}(\theta)$:
  \begin{align*}
      &\mathbb{E}_{p\sim \tau}\left[t(\theta,p)-\theta c(p)\right]\\
      \leq &\mathbb{E}_{p\sim \tau}\left[T(\theta)-\theta H_{\theta}(p)\right]\\
      = &T(\theta)-\theta H_{\theta}(\mathbb{E}_{p\sim \tau}\left[p\right]) \\
      = &T(\theta)-\theta H_{\theta}(\mathbb{E}_{p\sim \mathcal{X}(\theta)}\left[p\right])\\
      = &\mathbb{E}_{p\sim \mathcal{X}(\theta)}\left[t(\theta,p)-\theta c(p)\right],
  \end{align*}
\noindent where by construction we have $t(\theta,p)-\theta c(p)\leq T(\theta)-\theta H_{\theta}(p)$, $\forall p\in\Delta(\Omega)$, the first equality holds by $H_{\theta}$ being affine, the second equality holds by $\tau$ and $\mathcal{X}(\theta)$ both being Bayes-plausible, and the last equality holds because $H_{\theta}(p)=c(p)$, $\forall p\in \support (\mathcal{X}(\theta))$.
  
\textbf{Honesty} Truthful reporting is guaranteed by the incentive compatibility of $(\mathcal{X},T)$. Conditional on obedience, the inequality in (\ref{newic}) reduces to one of the incentive-compatible constraint for $(\mathcal{X},T)$, i.e., $T(\theta)-\theta\cdot\mathcal{C}(\mathcal{X}(\theta))\geq T(\theta^{\prime})-\theta\cdot\mathcal{C}(\mathcal{X}(\theta^{\prime}))$.

\begin{remark}
    Non-redundancy is required for this contingent payment function to work. If $\mathcal{X}(\theta)$ has affinely dependent support, the researcher reporting $\theta$ can deviate to an experiment whose support is a proper subset $\support (\mathcal{X}(\theta))$. Doing so will preserve the same payment but strictly lower the cost. Thus, it is profitable for the researcher.
\end{remark}
\end{proof}

Our analysis complements \cite{Y2022} by showing that it is possible to induce any desired experiment-payment pair using contingent payments on both the reported type and the realized posterior. Because type revelation gives more flexibility in designing incentives, Theorem \ref{theorem:1} speaks not just to the optimal methods-based contract but to all incentive-compatible methods-based contracts. The intuition is closely related to Proposition 1 in \cite{RS2017}: when the researcher has unlimited liability, a ``forcing'' contract can solve the hidden-action problem.

The following lemma allows us to apply Theorem \ref{theorem:1} to an optimal methods-based contract. Therefore, results-based contracting can achieve the same expected payoff for the principal as methods-based contracting. 

\begin{lemma}\label{lemma:2}
   There always exists an optimal methods-based contract $(\mathcal{X}^{*},T^{*})$ such that $\mathcal{X}^{*}$ is non-redundant.
\end{lemma}

\begin{proof}[Proof of Lemma \ref{lemma:2}]
    See the Appendix \ref{section:appA}.
\end{proof}

In concluding this section, we will discuss the applications of results-based payments with screening. The main takeaway is that screening helps provide learning incentives. A type report can be interpreted as a cost estimate. To put this in context, consider the process of applying for a grant. Applicants are typically required to submit not just a proposal that outlines their projects but also a budget proposal that breaks down all the anticipated expenses. The practice of screening, particularly in terms of cost assessment, has been gaining increasing traction within the field of medical research. While there is a recognized shortfall in transparency concerning the cost reporting of clinical trials, ongoing efforts are being made to enhance this practice. A notable example of these initiatives is led by Doctors Without Borders, also known as M{\'e}decins Sans Fronti{\`e}res \citep{BBM2020}.

\subsection{Results-based payments without screening}
 
 In this subsection, the principal is restricted to using payments that can only depend on the realized posterior. Following the terminology in \cite{Y2022}, we define a \emph{results-based contract} as a contingent payment function $t:\Delta\left(\Omega\right)\to \mathbb{R}$. We say that $(\mathcal{X},t)$ is incentive-compatible if
\begin{align}\label{newic2}
    \mathbb{E}_{p\sim \mathcal{X}(\theta)}[t(p)]-\theta \mathcal{C}(\mathcal{X}(\theta))\geq\mathbb{E}_{p\sim \tau}[t(p)]-\theta \mathcal{C}(\tau) \quad \forall \theta\in \Theta, \forall \tau\in X.
\end{align}

Because the principal cannot price discriminate according to different reports, a results-based contract must provide incentives for all researcher types.

We aim to investigate whether $(\mathcal{X}^{*},T^{*})$ can be implemented using a results-based contract. In the case of two states, \cite{Y2022} demonstrates that the optimal choice function $\mathcal{X}^{*}$ must be Blackwell-monotone. In such a case, a binary, Blackwell-monotone choice function is implementable by a results-based contract that induces the payment rule $T^{*}$. However, as shown in Example \ref{example:1}, this may not be the case when there are more than two states, as the optimal choice function need not be Blackwell-monotone. Thus, it is important to understand how Blackwell monotonicity relates to implementability in a broader context. 

We will introduce an additional assumption on $\mathcal{X}^{*}$ going forward to simplify the analysis. This assumption requires that every experiment has the same number of posteriors as $|\Omega|$. While this assumption is restrictive, it allows for a clear characterization of implementable experiment choice functions.\footnote{It rules out cases where the number of posteriors in the support of the desired experiment is strictly less than the number of states. Without this assumption, comprehending the outcome-equivalence result becomes challenging due to freedom in separating the adjacent types.} 

\begin{definition}
Experiment choice function $\mathcal{X}$ has \emph{full dimension}, if it is non-redundant and satisfies $|\support (\mathcal{X}(\theta))|=|\Omega|$, $\forall \theta\in \Theta$. 
\end{definition}

In particular, the optimal experiment choice function in Example \ref{example:1} has full dimension. The number of posteriors included in the support generally depends on the function $v(p)-g(\theta) c(p)$ and the prior. With Shannon entropy cost, \cite{CDL2019} provides a test for whether an action should be chosen at optimum, which can help determine the number of posteriors. While this assumption may not be applicable in all cases, it can be innocuous in some applications, such as when the principal faces a "matching-the-state" decision problem with a uniform prior. 

Before the main results, it will be convenient to introduce the following notations. We denote the support of $\mathcal{X}^{*}(\theta)$ by $S_{\theta}$. Slightly abusing notations, we will write $S_{\theta_{k}}$ as $S_{k}$ when we refer to an indexed type space. 

Under full dimensionality, $S_{k}$ is a basis for the affine subspace that contains $\Delta(\Omega)$. For any posterior in the simplex, there is a unique way to represent it as an affine combination of the elements in $S_{k}$. Write $S_{k}=\{p_{1}^{k}, p_{2}^{k}, \ldots, p_{|\Omega|}^{k}\}$. Formally, $\forall p\in \Delta(\Omega)$, $\exists$ unique $\alpha^{k}(p)\equiv (\alpha_{1}^{k},\alpha_{2}^{k},\ldots,\alpha_{|\Omega|}^{k})\in \mathbb{R}^{|\Omega|}$ such that $p=\sum_{i=1}^{|\Omega|}\alpha_{i}^{k}p_{i}^{k}$ and $\sum_{i=1}^{|\Omega|}\alpha_{i}^{k}=1$. Define $H_{k}(p)=\sum_{i=1}^{|\Omega|}\alpha_{i}^{k}(p)c(p_{i}^{k})$.\footnote{The definition here is identical to that of $H_{\theta}$ in the proof of Theorem \ref{theorem:1}. To simplify notation, we write $H_{k}$ instead of $H_{\theta_{k}}$.} This function $H_{k}$ is affine over $\Delta(\Omega)$ by definition and will play an important role in the following analysis. 

We introduce a new property of choice functions named \emph{strong $c$-monotonicity}. The definition involves an average of affine functions $H_{k}$ weighted by the relative distances between types. It requires that certain inequalities hold when we compare the cost function to this affine function over a particular set of posteriors. 

\begin{definition} An experiment choice function $\mathcal{X}$ is \emph{strongly $c$-monotone} if the following inequalities hold for all $i,j\in\{1,2,\ldots, N\}$ such that $i<j$.
\begin{align}
     c(p)\leq \sum_{k=i}^{j-1}\frac{\theta_{k}-\theta_{k+1}}{\theta_{i}-\theta_{j}}H_{k}(p) \text{, }\forall p\in S_{i} \text{ and } c(p) \geq \sum_{k=i}^{j-1}\frac{\theta_{k}-\theta_{k+1}}{\theta_{i}-\theta_{j}}H_{k}(p)   \text{, } \forall p\in S_{j} \label{eq:sm}
\end{align}
\end{definition}

The theorem below establishes an outcome equivalence between results-based contracting and methods-based contracting. If the principal wants to implement an experiment choice function with full dimension, the equivalence between two contracting regimes requires exactly strong $c$-monotonicity.

\begin{theorem}\label{theorem:2}
    Let $(\mathcal{X}^{*},T^{*})$ be an optimal methods-based contract and suppose that $\mathcal{X}^{*}$ is fully dimensional. There exists a contingent payment rule $t: \Delta(\Omega) \to \mathbb{R}$ such that $(\mathcal{X}^{*},t)$ is incentive-compatible and $\mathbb{E}_{p\sim \mathcal{X}^{*}(\theta)}[t(p)]= T^{*}(\theta)$, if and only if $\mathcal{X}^{*}$ is strongly $c$-monotone.
\end{theorem}
\begin{proof}[Proof of Theorem \ref{theorem:2}]
    See the Appendix \ref{section:appA}.
\end{proof}

With two states, it is known that the outcome-equivalence result relies on Blackwell monotonicity, a notion independent of the specific cost function. Theorem \ref{theorem:2} generalizes \citeauthor{Y2022}'s \citeyearpar{Y2022} insights to encompass settings with more than two states, pointing out that this result does depend on the cost function in general settings. Furthermore, we demonstrate that payments based solely on the realized posterior can be applied more broadly. In Example \ref{example:1}, we can establish that the optimal choice function is strongly $c$-monotone, even though it is not Blackwell-monotone. As per Theorem \ref{theorem:2}, results-based contracting attains the same optimal value for the principal as methods-based contracting. Later we will explain how to construct the optimal results-based contract. 

To help readers understand strong $c$-monotonicity, we have a detailed discussion in Section \ref{section:discussion}. In particular, symmetric settings often guarantee strong $c$-monotonicity, which carries significant implications for designing incentives.
	
\subsubsection{More on strong c-monotonicity}\label{section:discussion}

We first introduce two conditions that are closely related to strong $c$-monotonicity. These conditions can be useful in determining whether an experiment choice function is strongly $c$-monotone. 

\begin{proposition}\hfill
\label{proposition:2}
\begin{enumerate}
    \item A sufficient condition for strong $c$-monotonicity is for all $k\in\{1,2,\ldots, N-1\}$,
\begin{align}
     c(p)\leq H_{k}(p) \text{, if} ~p\in \cup_{i<k} S_{i} \text{; and } c(p) \geq H_{k}(p)   \text{, if} ~p\in \cup_{i>k} S_{i}. \label{eq:n}
\end{align}
    \item A necessary condition for strong $c$-monotonicity is for all $k\in\{1,2,\ldots, N-1\}$,
\begin{align}
     c(p)\leq H_{k}(p) \text{, if} ~ p\in S_{k-1} \text{; and } c(p) \geq H_{k}(p)   \text{, if} ~ p\in S_{k+1}. \quad(\text{Put } S_{0}=\emptyset. )\label{eq:nadj}
\end{align}
\end{enumerate}
\end{proposition}

\begin{proof}[Proof of Proposition \ref{proposition:2}]
    See the Appendix \ref{section:appA}.
\end{proof}

\begin{figure}[http]
	\centering
	\includegraphics[width=0.7\textwidth]{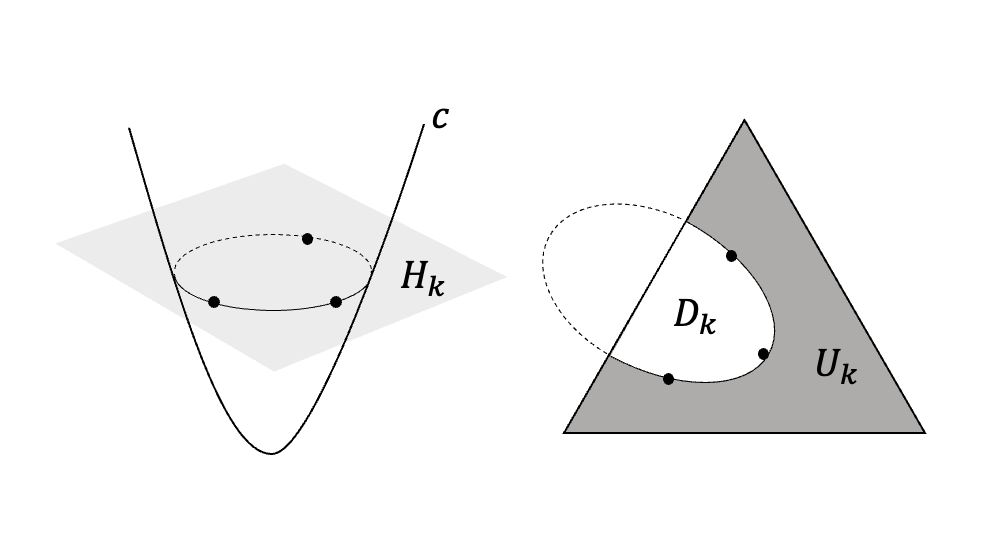}
	\caption{$U_{k}$ and $D_{k}$ in the simplex}
    \label{figure:2}
\end{figure}

To interpret, the idea behind these two conditions is that $\mathcal{X}$ has a nested structure with respect to $c$. Namely, for $k\in \{1,2,\ldots,N-1\}$, we can identify two sets in the simplex: $U_{k}=\{p\in\Delta(\Omega)|c(p)\geq H_{k}(p)\}$ and $D_{k}=\{p\in\Delta(\Omega)|c(p) \leq H_{k}(p)\}$. Geometrically, $D_{k}$ is a convex set that inscribes the convex hull of $S_{k}$ and $U_{k}$ is the closure of $D_{k}$'s complement. As shown in Figure \ref{figure:2}, it is easy to visualize these sets for the case with three states. 

Condition (\ref{eq:n}) means for any fixed $k$, if $i<k$, then $S_{i}\subseteq D_{k}$ and if $i>k$, then $S_{i}\subseteq U_{k}$. Condition (\ref{eq:nadj}) only has the nested requirement for adjacent types, meaning $S_{k-1}\subseteq D_{k}$ and $S_{k+1}\subseteq U_{k}$. Note that we only need $N-1$ affine hyperplanes $H_{k}$ here.\footnote{We may define a partial order on $X$ and rewrite condition (\ref{eq:n}) and (\ref{eq:nadj}) in terms of the order. But we have to treat the most efficient type $\theta_{N}$ separately because we don't need $c(p)\leq H_{N}(p)$ for $p\in S_{i}$ such that $i<N$.} The sufficient condition consists of $(N-1)^{2}$ inequalities, while the necessary condition only has $2(N-1)-1$. In the special case with two types ($N=2$), they have the same content, so both are equivalent to strong $c$-monotonicity. 

\begin{corollary} \label{corollary:1} With $N=2$, conditions (\ref{eq:sm}), (\ref{eq:n}) and (\ref{eq:nadj}) are equivalent.
\end{corollary}

Next, we compare strong $c$-monotonicity to the previously mentioned concepts: $c$-monotonicity and Blackwell monotonicity. 

\begin{proposition}\label{proposition:3} The following statements hold: 
\begin{enumerate}
    \item Strong $c$-monotonicity implies $c$-monotonicity; the converse is not true. 
    \item With $|\Omega|=2$, strong $c$-monotonicity is equivalent to Blackwell monotonicity.
    \item With $|\Omega|>2$, Blackwell monotonicity does not imply strong $c$-monotonicity; strong $c$-monotonicity does not imply Blackwell monotonicity.
\end{enumerate}
\end{proposition}
\begin{proof}[Proof of Proposition \ref{proposition:3}]
    See the Appendix \ref{section:appA}.
\end{proof}

The first statement in Proposition \ref{proposition:2} motivates us to use the name of strong $c$-monotonicity. The relationship between strong $c$-monotonicity and Blackwell monotonicity crucially depends on the dimension of state space. The reason is that with $|\Omega|=2$, the convex hull of $S_{k}$ conincides with $D_{k}$, but with $|\Omega|>2$, the convex hull of $S_{k}$ is generally a proper subset of $D_{k}$.

The next Proposition \ref{proposition:4} strengthens $c$-monotonicity to strong $c$-monotonicity. 

\begin{definition}
An experiment $\tau$ is symmetric if there exists $h \in \mathbb{R}$ such that for every $p\in \support (\tau)$, $c(p)=h$. We say that an experiment choice function $\mathcal{X}$ is symmetric if for every $\theta \in \Theta$, $\mathcal{X}(\theta)$ is symmetric.
\end{definition}
\begin{proposition} \label{proposition:4} If an experiment choice function $\mathcal{X}$ is $c$-monotone and symmetric, then it is strongly $c$-monotone.
\end{proposition}
\begin{proof}[Proof of Proposition \ref{proposition:4}]
If $\mathcal{X}$ is symmetric, then for every $k$, $H_{k}$ is a constant function, i.e., $\exists h_{k}\in \mathbb{R}$: $\forall p \in \Delta(\Omega)$, $H_{k}(p)=h_{k}$. Thus, $D_{k}$ is the sublevel set of function $c$ where it takes on the constant value of $h_{k}$. Fix $k \in \{1,2,\ldots,N-1\}$. First, suppose $i<k$. By $c$-monotonicity, we have $h_{i}<h_{k}$. For any $p\in S_{i}$, $c(p)=h_{i}<h_{k}$, which means $p\in D_{k}$. Next, suppose $i>k$. $c$-monotonicity implies $h_{i}>h_{k}$. For any $p\in S_{i}$, $c(p)=h_{i}>h_{k}$, which means $p\in U_{k}$. Condition (\ref{eq:n}) holds and so does strong $c$-monotonicity.
\end{proof}

 Recall that the optimal experiment choice function is always $c$-monotone. If it is symmetric, then the optimal methods-based contract can be replicated by a results-based contract. We also know that the challenge with providing learning incentives must come from ``asymmetry''. In a pure moral hazard model, \cite{RS2017} shows that an experiment is implementable if and only if it is symmetric.  One may ask whether symmetry is necessary for strong $c$-monotonicity in our setting. The answer is no. We can refer back to the optimal experiment choice function in Example \ref{example:1}, which satisfies strong $c$-monotonicity but not symmetry. Therefore, it is possible to obtain the equivalence result even when the principal seeks to implement an experiment choice function that forgoes symmetry.

\subsubsection{An illustration of our results with Example \ref{example:1}}

We will illustrate an application of Theorem \ref{theorem:2} to Example \ref{example:1} by showing that the optimal experiment choice function is strongly $c$-monotone and then discussing how to construct an outcome-equivalent results-based contract.

 \addtocounter{example}{-1}
\begin{example} (Continued.)
 
 With two types, strong $c$-monotonicity reduces to a simple condition $S_{2}\subseteq U_{1}$. As shown in the Figure \ref{figure:3}, the vertices of the thick-lined triangle represent posteriors in $S_{1}=\{p_{1}, p_{1}^{\prime}, p_{1}^{\prime\prime}\}$ for the less efficient type $\theta_{1}$; the vertices of the dashed-lined triangle are posteriors in $S_{2}=\{p_{2}, p_{2}^{\prime}, p_{2}^{\prime\prime}\}$ for the more efficient type $\theta_{2}$. The light grey area represents set $D_{1}$; the dark grey area represents set $U_{1}$. We find $S_{2} \subseteq U_{1}$.

\begin{figure}[http]
	\centering
	\includegraphics[width=0.7\textwidth]{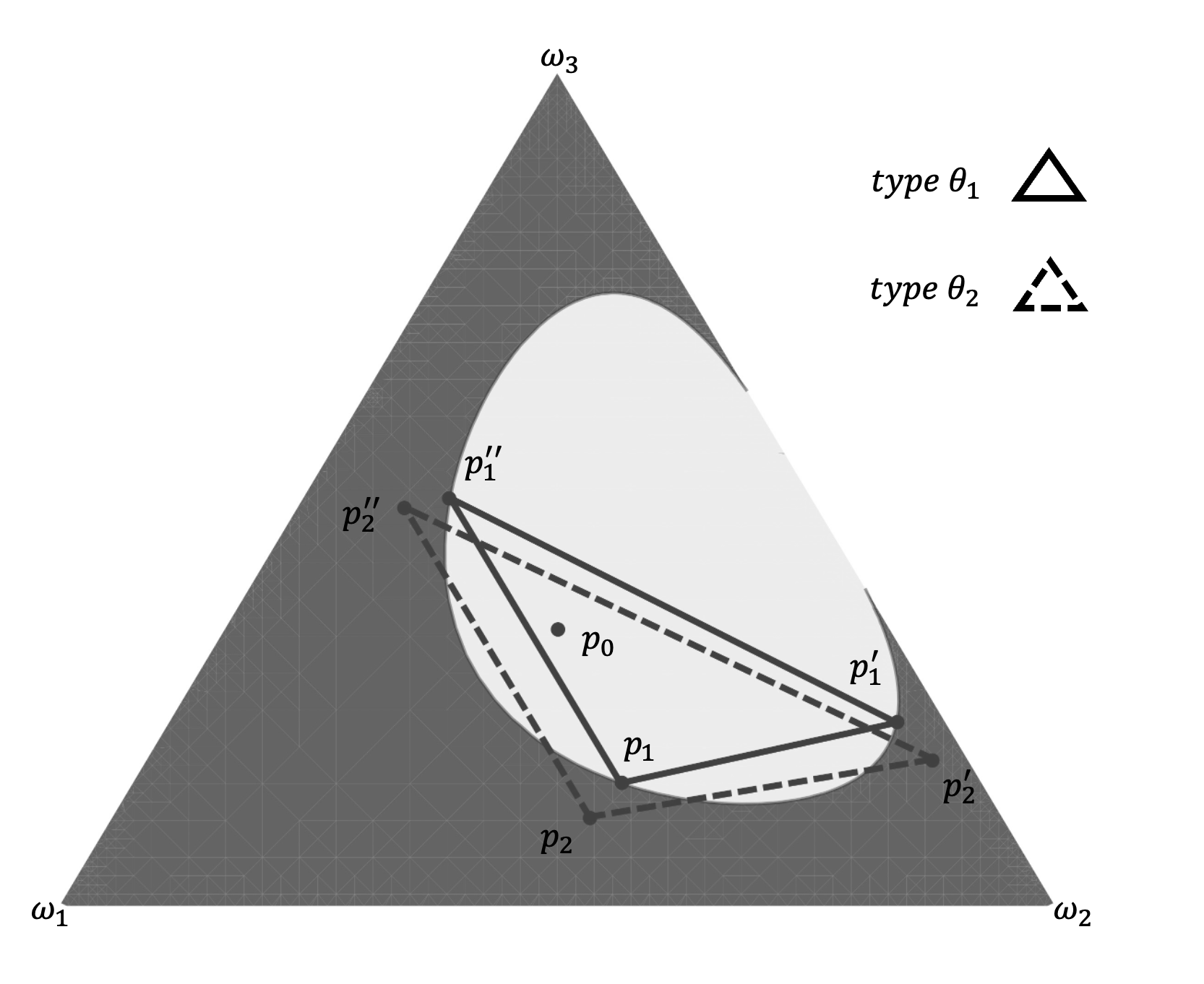}
	\caption{The optimal experiment choice function is strongly $c$-monotone}
	\label{figure:3}
\end{figure}

	\end{example}

We can pick a sufficiently small $\underline{t}\in \mathbb{R}$ and construct a results-based contract:
\begin{align*}
t(p)=\begin{cases}
    \theta_{1}c(p)\text{, if } p\in S_{1}\\
    \theta_{2}c(p)+(\theta_{1}-\theta_{2})H_{1}(p)\text{, if } p\in S_{2}\\
    \underline{t}\text{, otherwise}
    \end{cases}
\end{align*}

It is easy to check that $(\mathcal{X}^{*},t)$ is incentive-compatible and  $\mathbb{E}_{p\sim \mathcal{X}^{*}(\theta)}[t(p)]= T^{*}(\theta)$. For each type of researcher, we can draw his payoff as a function of the realized posterior. Under strong $c$-monotonicity, it follows that there is a hyperplane weakly above all the points. We can identify this hyperplane by an affine function $s_{k,t}:\Delta(\Omega)\to \mathbb{R}$. The type-$\theta_{k}$ researcher's highest expected payoff is $s_{k,t}(p_{0})$, achieved by choosing $\mathcal{X}^{*}(\theta_{k})$. Moreover, according to (\ref{eq:Tstar1}) and (\ref{eq:Tstar2}), simple calculation can confirm that $s_{k,t}(p_{0})=T^{*}(\theta_{k})-\theta_{k} \mathcal{C}(\mathcal{X}^{*}(\theta_{k}))$.

Strong $c$-monotonicity is important for the type-$\theta_{1}$ researcher not to mimic type-$\theta_{2}$. Suppose that it is violated. The contract $t$ we constructed won't be incentive-compatible. For example, if $c(p_{2})<H_{1}(p_{2})$, then under contract $t$, the type-$\theta_{1}$ researcher's payoff at $p_{2}$ will be above $s_{1}(p_{2})$. The researcher can get a strictly higher expected payoff by choosing an experiment with support $\{p_{2},p_{1},p_{1}^{\prime},p_{1}^{\prime\prime}\}$ than that of choosing $\mathcal{X}^{*}(\theta_{1})$. Interestingly, we can strengthen this result, as the failure of strong $c$-monotonicity will prevent us from getting any incentive-compatible results-based contract that induces $(\mathcal{X}^{*},T^{*})$. In other words, strong $c$-monotonicity is necessary for implementing the optimal methods-based contract. 

\newpage
\nocite{*}
\bibliography{CfHR}

\begin{thebibliography}{31}
\newcommand{\enquote}[1]{``#1''}
\expandafter\ifx\csname natexlab\endcsname\relax\def\natexlab#1{#1}\fi

\bibitem[\protect\citeauthoryear{Aumann and Maschler}{Aumann and
  Maschler}{1995}]{AM1995}
\textsc{Aumann, R.~J. and M.~Maschler} (1995): \enquote{Repeated games with
  incomplete information, with the collaboration of {R}ichard {S}tearns,} .

\bibitem[\protect\citeauthoryear{Aybas and Turkel}{Aybas and
  Turkel}{2022}]{AT2022}
\textsc{Aybas, Y.~C. and E.~Turkel} (2022): \enquote{Persuasion with Coarse
  Communication,} Working paper, Stanford University.

\bibitem[\protect\citeauthoryear{Azrieli}{Azrieli}{2021}]{A2021}
\textsc{Azrieli, Y.} (2021): \enquote{Monitoring experts,} \emph{Theoretical
  Economics}, 16, 1313--1350.

\bibitem[\protect\citeauthoryear{Azrieli}{Azrieli}{2022}]{A2022}
---\hspace{-.1pt}---\hspace{-.1pt}--- (2022): \enquote{Delegated expertise:
  Implementability with peer-monitoring,} \emph{Games and Economic Behavior},
  132, 240--254.

\bibitem[\protect\citeauthoryear{Barel, Boman, and Morten}{Barel
  et~al.}{2020}]{BBM2020}
\textsc{Barel, A., L.~Boman, and C.~Morten} (2020): \enquote{Clinical Trial
  Cost Transparency at the National Institutes of Health: Law and Policy
  Recommendations,} Tech. rep., Engelberg Center on Innovation Law and Policy,
  NYU School of Law, https://ssrn.com/abstract=3880756.

\bibitem[\protect\citeauthoryear{Bergemann and Bonatti}{Bergemann and
  Bonatti}{2019}]{BB2019}
\textsc{Bergemann, D. and A.~Bonatti} (2019): \enquote{Markets for Information:
  An Introduction,} \emph{Annual Review of Economics}, 11, 85--107.

\bibitem[\protect\citeauthoryear{Bergemann, Bonatti, and Smolin}{Bergemann
  et~al.}{2018}]{BBS2018}
\textsc{Bergemann, D., A.~Bonatti, and A.~Smolin} (2018): \enquote{The design
  and price of information,} \emph{American Economic Review}, 108, 1--48.

\bibitem[\protect\citeauthoryear{Bergemann and Morris}{Bergemann and
  Morris}{2019}]{BM2019}
\textsc{Bergemann, D. and S.~Morris} (2019): \enquote{Information design: A
  unified perspective,} \emph{Journal of Economic Literature}, 57, 44--95.

\bibitem[\protect\citeauthoryear{Blackwell}{Blackwell}{1951}]{B1951}
\textsc{Blackwell, D.} (1951): \enquote{Comparison of Experiments,} in
  \emph{Proceedings of the Second Berkeley Symposium on Mathematical Statistics
  and Probability}, University of California Press, vol.~2, 93--102.

\bibitem[\protect\citeauthoryear{Blackwell}{Blackwell}{1953}]{B1953}
---\hspace{-.1pt}---\hspace{-.1pt}--- (1953): \enquote{Equivalent Comparisons
  of Experiments,} \emph{The Annals of Mathematical Statistics}, 24, 265--272.

\bibitem[\protect\citeauthoryear{Caplin, Dean, and Leahy}{Caplin
  et~al.}{2017}]{CDL2017}
\textsc{Caplin, A., M.~Dean, and J.~Leahy} (2017): \enquote{Rationally
  inattentive behavior: characterizing and generalizing {S}hannon entropy,}
  Working paper, National Bureau of Economic Research.

\bibitem[\protect\citeauthoryear{Caplin, Dean, and Leahy}{Caplin
  et~al.}{2019}]{CDL2019}
---\hspace{-.1pt}---\hspace{-.1pt}--- (2019): \enquote{Rational inattention,
  optimal consideration sets, and stochastic choice,} \emph{The Review of
  Economic Studies}, 86, 1061--1094.

\bibitem[\protect\citeauthoryear{Carroll}{Carroll}{2019}]{C2019}
\textsc{Carroll, G.} (2019): \enquote{Robust incentives for information
  acquisition,} \emph{Journal of Economic Theory}, 181, 382--420.

\bibitem[\protect\citeauthoryear{Clark and Reggiani}{Clark and
  Reggiani}{2021}]{CR2021}
\textsc{Clark, A. and G.~Reggiani} (2021): \enquote{Contracts for acquiring
  information,} Working paper, https://arxiv.org/abs/2103.03911.

\bibitem[\protect\citeauthoryear{Denti}{Denti}{2022}]{D2022}
\textsc{Denti, T.} (2022): \enquote{Posterior separable cost of information,}
  \emph{American Economic Review}, 112, 3215--59.

\bibitem[\protect\citeauthoryear{H{\"a}fner and Taylor}{H{\"a}fner and
  Taylor}{2018}]{HT2018}
\textsc{H{\"a}fner, S. and C.~R. Taylor} (2018): \enquote{Contracting for
  research: Moral hazard and the incentive to overstate significance,} Working
  paper, https://ssrn.com/abstract=3229927.

\bibitem[\protect\citeauthoryear{Kamenica}{Kamenica}{2019}]{K2019}
\textsc{Kamenica, E.} (2019): \enquote{Bayesian persuasion and information
  design,} \emph{Annual Review of Economics}, 11, 249--272.

\bibitem[\protect\citeauthoryear{Kamenica and Gentzkow}{Kamenica and
  Gentzkow}{2011}]{KG2011}
\textsc{Kamenica, E. and M.~Gentzkow} (2011): \enquote{Bayesian persuasion,}
  \emph{American Economic Review}, 101, 2590--2615.

\bibitem[\protect\citeauthoryear{Li}{Li}{2021}]{L2021}
\textsc{Li, Y.} (2021): \enquote{Selling data to an agent with endogenous
  information,} Working paper, Northwestern University.

\bibitem[\protect\citeauthoryear{Maskin and Riley}{Maskin and
  Riley}{1984}]{MR1984}
\textsc{Maskin, E. and J.~Riley} (1984): \enquote{Monopoly with incomplete
  information,} \emph{The RAND Journal of Economics}, 15, 171--196.

\bibitem[\protect\citeauthoryear{Mat{\v{e}}jka and McKay}{Mat{\v{e}}jka and
  McKay}{2015}]{MM2015}
\textsc{Mat{\v{e}}jka, F. and A.~McKay} (2015): \enquote{Rational inattention
  to discrete choices: A new foundation for the multinomial logit model,}
  \emph{American Economic Review}, 105, 272--98.

\bibitem[\protect\citeauthoryear{Min}{Min}{2021}]{M2021}
\textsc{Min, D.} (2021): \enquote{Screening for Experiments,} Working paper,
  New York University Abu Dhabi.

\bibitem[\protect\citeauthoryear{Myerson}{Myerson}{1981}]{M1981}
\textsc{Myerson, R.~B.} (1981): \enquote{Optimal auction design,}
  \emph{Mathematics of Operations Research}, 6, 58--73.

\bibitem[\protect\citeauthoryear{Myerson}{Myerson}{1982}]{M1982}
---\hspace{-.1pt}---\hspace{-.1pt}--- (1982): \enquote{Optimal coordination
  mechanisms in generalized principal–agent problems,} \emph{Journal of
  Mathematical Economics}, 10, 67--81.

\bibitem[\protect\citeauthoryear{Osband}{Osband}{1989}]{O1989}
\textsc{Osband, K.} (1989): \enquote{Optimal forecasting incentives,}
  \emph{Journal of Political Economy}, 97, 1091--1112.

\bibitem[\protect\citeauthoryear{Rappoport and Somma}{Rappoport and
  Somma}{2017}]{RS2017}
\textsc{Rappoport, D. and V.~Somma} (2017): \enquote{Incentivizing information
  design,} Working paper, Columbia University.

\bibitem[\protect\citeauthoryear{Whitmeyer and Zhang}{Whitmeyer and
  Zhang}{2022}]{WZ2022}
\textsc{Whitmeyer, M. and K.~Zhang} (2022): \enquote{Buying Opinions,} Working
  paper, Arizona State University.

\bibitem[\protect\citeauthoryear{Wu}{Wu}{2018}]{W2018}
\textsc{Wu, W.} (2018): \enquote{Sequential Bayesian Persuasion,} Working
  paper, University of Arizona.

\bibitem[\protect\citeauthoryear{Yang}{Yang}{2022}]{YA2022}
\textsc{Yang, K.~H.} (2022): \enquote{Selling consumer data for profit: Optimal
  market-segmentation design and its consequences,} \emph{American Economic
  Review}, 112, 1364--1393.

\bibitem[\protect\citeauthoryear{Yoder}{Yoder}{2022}]{Y2022}
\textsc{Yoder, N.} (2022): \enquote{Designing incentives for heterogeneous
  researchers,} \emph{Journal of Political Economy}, 130, 2018--2054.

\bibitem[\protect\citeauthoryear{Zerme{\~{n}}o}{Zerme{\~{n}}o}{2011}]{Z2011}
\textsc{Zerme{\~{n}}o, L.} (2011): \enquote{A principal-expert model and the
  value of menus,} Working paper, Massachusetts Institute of Technology.

\end{thebibliography}

\newpage
\appendix

\section{Proofs Omitted from Main Text} \label{section:appA}

\begin{proof}[Proof of violation of Blackwell monotonicity in Example \ref{example:1}.]\hfill

Define six posterior beliefs as follows.  
\begin{align*}
p_{1}=(0.3626,0.4899,0.1475)\\
p_{1}^{\prime}=(0.0491,0.7308,0.2201)\\
p_{1}^{\prime\prime}=(0.3626,0.1475,0.4899)\\
p_{2}=(0.4141,0.4790,0.1069)\\
p_{2}^{\prime}=(0.0340,0.7898,0.1762)\\
p_{2}^{\prime\prime}=(0.4141,0.1069,0.4790)
\end{align*}
    
We solved the optimal experiment choice function following Proposition 2 of \cite{CDL2019}. $\mathcal{X}^{*}(\theta_{1})$ is a distribution over posteriors such that $p_{1}$ occurs with probability $0.3838$, $p_{1}^{\prime}$ occurs with probability $0.0933$ and $p_{1}^{\prime\prime}$ occurs with probability $0.5229$. $\mathcal{X}^{*}(\theta_{2})$ is a distribution over posteriors such that $p_{2}$ occurs with probability $0.2186$, $p_{2}^{\prime}$ occurs with probability $0.2125$ and $p_{2}^{\prime\prime}$ occurs with probability $0.5689$. 

Note that $p_{1}^{\prime} \in \support{\left(\mathcal{X}^{*}\left(\theta_{1}\right)\right)}$ and $p_{1}^{\prime} \not\in \convexhull(\support{\left(\mathcal{X}^{*}\left(\theta_{2}\right)\right)})$, which means $\mathcal{X}^{*}(\theta_{2})$ is not Blackwell more informative than $\mathcal{X}^{*}(\theta_{1})$.
\end{proof}

\begin{proof}[Proof of Lemma \ref{lemma:2}]
We will show that for every $\theta$, there exists an experiment $\mathcal{X}^{*}(\theta)\in X$ that solves the problem (\ref{eq:type-specific}) and has affinely independent support. 

    Notice that each type-specific problem in (\ref{eq:type-specific}) has the same structure as the sender's problem in a \emph{Bayesian persuasion} model. It is known that solutions exist and can be found using the concavification approach \citep{KG2011}. 
    
    Suppose that a solution to problem (\ref{eq:type-specific}) has affinely dependent support. We can form a new experiment by dropping some redundant posterior that still preserves the optimality. The proof follows directly from \cite{AT2022}. 
\end{proof}

\textbf{Characterizing incentive-compatible results-based contracts}

\noindent Before the proof of Theorem \ref{theorem:2}, we want to introduce a necessary and sufficient secant hyperplane condition for a results-based contract to be incentive-compatible. This is a generalization of Proposition 2 in \cite{Y2022}. 

Suppose that $\mathcal{X}^{*}$ is fully dimensional. Recall that we denote the barycentric coordinates of $p$ over $S_{k}$ as $\alpha^{k}(p)$. Let $u_{k,t}(p)= t(p)-\theta_{k} c(p)$ be the type-$\theta_{k}$ researcher's payoff when holding belief $p$. For the following proofs, define $s_{k,t}(p)=\sum_{i=1}^{|\Omega|}\alpha_{i}^{k} (p)\left[u_{k, t}(p_{i}^{k})\right]$. By definition, $s_{k,t}$ represents the affine hyperplane uniquely determined by $|\Omega|$ points from $\{\left(p,u_{k, t}\left(p\right)\right): p\in S_{k}\}$ and $s_{k,t}$ equals $u_{k,t}$ at every $p\in S_{k}$.

\begin{lemma}\label{lemma:4}
Suppose that $\mathcal{X}^{*}$ has full dimension. A results-based contract $(\mathcal{X}^{*},t)$ is incentive-compatible if and only if the secant hyperplane condition holds: 
\begin{align} \label{eq:shc}
	    s_{k,t}(p) \geq u_{k,t}(p) \quad \forall p\in \Delta(\Omega) \text{ and }\forall k \in \{1,2,\ldots, N\}
\end{align}
\end{lemma}

 \begin{proof}[Proof of Lemma \ref{lemma:4}]\hfill

    ($\implies$) We will prove this direction by contradiction. Suppose that the secant hyperplane condition (\ref{eq:shc}) is not true. There must exist $q\in \Delta(\Omega)$ such that $s_{k,t}(q) < u_{k,t}(q)$ for some $k$. Let $|\Omega|=n$. Under full dimensionality, write $S_{k}=\{p^{1},p^{2},\ldots, p^{n}\}$. We want to show that replacing a posterior in $S_{k}$ by $q$ can form a Bayes-plausible information structure $\tau$, then $\tau$ is a profitable deviation for the type-$\theta_{k}$ researcher. 

    By Bayes plausibility of $\mathcal{X}^{*}(\theta_{k})$, we can find $\lambda=(\lambda_{1},\ldots, \lambda_{n})\in \mathbb{R}_{++}^{n}$ with $\sum_{i=1}^{n}\lambda_{i}=1$ such that $p_{0}=\sum_{i=1}^{n}\lambda_{i}p^{i}$. 
    
    Because $S_{k}$ is affinely independent, we can find $\mu=(\mu_{1},\ldots, \mu_{n})\in \mathbb{R}^{n}$ with $\sum_{i=1}^{n}\mu_{i}=1$ such that $q=\sum_{i=1}^{n}\mu_{i}p^{i}$. 
    
    Note that there is $\mu_{i}>0$ for some $i$, otherwise we cannot have $\sum_{i=1}^{n}\mu_{i}=1$. 
    
    Denote $j\in \arg\min\{\frac{\lambda_{i}}{\mu_{i}}: \mu_{i}>0\}$. Because $\mu_{j}>0$, we can write $p^{j}$ as an affine combination using $q$ and other elements in $S_{k}$. Thus, $p^{j}=\frac{1}{\mu_{j}}\left(q-\sum_{i\neq j}\mu_{i}p^{i}\right)$. 
    
    Substituting $p^{j}$, we write  $p_{0}=\sum_{i=1}^{n}\lambda_{i}p^{i}$ as
    \begin{align*}
        p_{0}=\frac{\lambda_{j}}{\mu_{j}}q+\sum_{i\neq j}\left(\lambda_{i}-\frac{\mu_{i}}{\mu_{j}}\lambda_{j}\right)p^{i}.
    \end{align*}

We will verify that $p_{0}$ is written as a convex combination of points in $\{q\}\cup \left(S_{k}\setminus \{p^{j}\}\right)$. Recall that $\lambda_{i}>0$, $\forall i$ and $\mu_{j}>0$. It is trivial that $\frac{\lambda_{j}}{\mu_{j}}>0$. For all $i$ with $\mu_{i}<0$, 
$\lambda_{i}-\frac{\mu_{i}}{\mu_{j}}\lambda_{j}>0$; for all $i\neq j$ with $\mu_{i}>0$, $\lambda_{i}-\frac{\mu_{i}}{\mu_{j}}\lambda_{j}\geq \lambda_{i}-\frac{\mu_{i}}{\mu_{i}}\lambda_{i}=0$. Next, summing them up, we have $\frac{\lambda_{j}}{\mu_{j}}+\sum_{i\neq j}\left(\lambda_{i}-\frac{\mu_{i}}{\mu_{j}}\lambda_{j}\right)= \lambda_{j}+\sum_{i\neq j}\lambda_{i}=1$.

Define a distribution over posteriors $\tau$ such that $q$ occurs with probability $\frac{\lambda_{j}}{\mu_{j}}$ and for every $i\neq j$, $p^{i}$ occurs with probability $\lambda_{i}-\frac{\mu_{i}}{\mu_{j}}\lambda_{j}$.

We show that the type-$\theta$ researcher gets strictly higher expected payoff by choosing $\tau$ instead of $\mathcal{X}^{*}(\theta_{k})$:
    \begin{align*}
        \mathbb{E}_{p\sim \tau}\left[u_{k,t}(p)\right]&=\tau(q) u_{k,t}(q)+\sum_{i\neq j}\tau(p^{i}) s_{k,t}(p^{i})\\
        &>\tau(q) s_{k,t}(q)+\sum_{i\neq j}\tau(p^{i}) s_{k,t}(p^{i})\\
        &=\mathbb{E}_{p\sim \tau}\left[s_{k,t}(p)\right]\\
        &=s_{k,t}(\mathbb{E}_{p\sim \tau}\left[p\right])\\
        &=s_{k,t}(\mathbb{E}_{p\sim \mathcal{X}^{*}(\theta_{k})}\left[p\right])\\
        &=\mathbb{E}_{p\sim \mathcal{X}^{*}(\theta_{k})}\left[s_{k,t}(p)\right]\\
        &=\mathbb{E}_{p\sim \mathcal{X}^{*}(\theta_{k})}\left[u_{k,t}(p)\right],
    \end{align*}
    \noindent where the strict inequality holds by assumption that $s_{k,t}(q) < u_{k,t}(q)$; the first and last equalities follow from the definition of $s_{\theta,t}$, which satisfies $s_{k,t}(p)=u_{k,t}(p)$ for all $p\in S_{k}$; the third equality is true because $s_{k,t}(p)$ is an affine function; and the fourth equality follows from Bayes plausibility, which means $\mathbb{E}_{p\sim \tau}\left[p\right]=\mathbb{E}_{p\sim \mathcal{X}^{*}(\theta_{k})}\left[p\right]=p_{0}$.

  This contradicts with incentive compatibility for the type-$\theta$ researcher.

($\impliedby$) Fix $\theta_{k}$ and take any experiment $\tau$. The researcher's payoff of choosing $\tau$ is bounded from the above by $s_{k,t}(p_{0})$, because
    $$\mathbb{E}_{p\sim \tau}\left[u_{k,t}(p)\right]\leq\mathbb{E}_{p\sim\tau}\left[s_{k,t}(p)\right]=s_{k,t}(\mathbb{E}_{p\sim \tau}\left[p\right])=s_{k,t}(p_{0}),$$
    \noindent where the inequality follows from condition (\ref{eq:shc}), the first equality holds due to  
    $s_{k,t}$ being affine and the second equality uses Bayes plausibility.
    
    Notice that $\mathcal{X}^{*}(\theta_{k})$ attains the payoff $s_{k,t}(p_{0})$, so it is optimal for the type-$\theta_{k}$ researcher. This proves condition (\ref{newic2}), so $(\mathcal{X}^{*},t)$ is incentive-compatible. 
\end{proof}

\begin{proof}[Proof of Theorem \ref{theorem:2}]\hfill
 
 ($\impliedby$) Fix a sufficiently small $\underline{t}\in \mathbb{R}$. We can define a contingent payment rule $t^{*}:\Delta(\Omega)\to \mathbb{R}$ as follows.
 \begin{align*}
     t^{*}(p)=\begin{cases}
     \theta_{1}c(p) \quad &\text{for } p\in S_{1}\\
     \theta_{i}c(p)+\sum_{k=1}^{i-1}(\theta_{k}-\theta_{k+1})H_{k}(p) \quad &\text{for } p\in S_{i} \text{ and } i \in \{2,3,\ldots,N\}\\
     \underline{t} \quad &\text{otherwise} 
     \end{cases}
 \end{align*}

  For every $k\in \{1,2,\ldots,N\}$, we solve the affine function  $s_{k,t^{*}}(p)$ associated with $t^{*}$:
  \begin{align*}
   s_{1,t^{*}}(p)&=0\\
     s_{j,t^{*}}(p)&=\sum_{k=1}^{j-1}(\theta_{k}-\theta_{k+1})H_{k}(p) \quad \forall j \in\{2,3,\ldots,N\}
 \end{align*}
 
 Notice that by definition, every $s_{k,t^{*}}$ is continuous and does not depend on $\underline{t}$. Set $\underline{t}=\min_{q\in \Delta(\Omega)}\min_{k\in \mathbb{N}, 1\leq k\leq N}\{s_{k,t^{*}}(q)+\theta_{k}c(q)\}$. $\underline{t}$ is well-defined, because the minimum of a finite number of continuous functions is continuous and $\Delta(\Omega)$ is compact.
 
 Suppose that $\mathcal{X}^{*}$ is strongly $c$-monotone. We need to verify two statements: \textbf{(i)} the results-based contract $(\mathcal{X}^{*},t^{*})$ is incentive-compatible, and \textbf{(ii)} $t^{*}$ induces $T^{*}$.

    \textbf{(i)} In order to show that $(\mathcal{X}^{*},t^{*})$ is incentive-compatible, we will prove that $(\mathcal{X}^{*},t^{*})$ satisfies the secant hyperplane condition (\ref{eq:shc}) in Lemma \ref{lemma:4}. 
    
 We will verify that at every belief $p\in\Delta(\Omega)$, $s_{k,t^{*}}(p)\geq u_{k,t^{*}}(p)$ for all $k$.
 
 $\bullet$ Suppose $p\not\in \cup_{k=1}^{N}S_{k}$. In this case,  $t^{*}(p)=\underline{t}$. By definition, $\underline{t} \leq s_{k,t^{*}}(p) +\theta_{k} c(p)$ holds for all $k$. We have $s_{k,t^{*}}(p)\geq t^{*}(p)-\theta_{k} c(p)\equiv u_{k,t^{*}}(p)$, $\forall k$.
 
 $\bullet$ Suppose $p\in \cup_{k=1}^{N}S_{k}$. We first calculate the difference: 
 \begin{align*}
      &s_{1,t^{*}}(p) - u_{1,t^{*}}(p) \\
      =&\begin{cases}
     0 \quad &\text{for } p\in S_{1}\\
     \left(\theta_{1}-\theta_{j}\right)c(p)-\sum_{k=1}^{j-1}(\theta_{k}-\theta_{k+1})H_{k}(p) \quad &\text{for } p\in S_{j} \text{ and } j \geq 2
     \end{cases}
 \end{align*}
which is non-negative because by strong $c$-monotonicity, $c(p)\geq \sum_{k=1}^{j-1}\frac{\theta_{k}-\theta_{k+1}}{\theta_{1}-\theta_{j}} H_{k}(p)$ holds for any $p\in S_{j}$ and $j \geq 2$. Thus, we have $s_{1,t^{*}}(p) \geq u_{1,t^{*}}(p)$.

For $j\geq 2$, we also find the difference is non-negative:
  \begin{align*}
      &s_{j,t^{*}}(p) - u_{j,t^{*}}(p) \\
      =&\begin{cases}
     \sum_{k=i}^{j-1}(\theta_{k}-\theta_{k+1})H_{k}(p)-\left(\theta_{i}-\theta_{j}\right)c(p) \quad &\text{for } p\in S_{i} \text{ and all } i \text{ s.t. } 1 \leq i < j-1\\
     0 \quad &\text{for } p\in S_{j-1}\\
     0 \quad &\text{for } p \in S_{j}\\
     \left(\theta_{j}-\theta_{i}\right)c(p)-\sum_{k=j}^{i-1}(\theta_{k}-\theta_{k+1})H_{k}(p) \quad &\text{for } p\in S_{i} \text{ and all } i > j
     \end{cases}
 \end{align*}
 
  By strong $c$-monotonicity, $\sum_{k=i}^{j-1}(\theta_{k}-\theta_{k+1})H_{k}(p)-\left(\theta_{i}-\theta_{j}\right)c(p)$, $\forall p\in S_{i}$ and $\forall i$ s.t. $1 \leq i< j-1$; $\left(\theta_{j}-\theta_{i}\right)c(p)-\sum_{k=j}^{i-1}(\theta_{k}-\theta_{k+1})H_{k}(p)$ and $\forall i$ s.t. $i> j-1$. Therefore, we have $s_{j,t^{*}}(p) \geq u_{j,t^{*}}(p)$, $\forall j\geq 2$.
 
 \textbf{(ii)} Next, we want to show that $t^{*}$ induces the payment function $T^{*}$.
 
 \begin{itemize}
     \item[For $i=1$:]  $$\mathbb{E}_{p\sim \mathcal{X}^{*}\left(\theta_{1}\right)}\left[t^{*}\left(p\right)\right]=\mathbb{E}_{p\sim \mathcal{X}^{*}\left(\theta_{1}\right)}\left[\theta_{1}c(p)\right]=T^{*}(\theta_{1})$$
     \item[For $i\geq 2$:] \begin{align*}
     \mathbb{E}_{p\sim \mathcal{X}^{*}\left(\theta_{i}\right)}\left[t^{*}\left(p\right)\right]&=\mathbb{E}_{p\sim \mathcal{X}^{*}\left(\theta_{i}\right)}\left[ \theta_{i}c(p)+\sum_{k=1}^{i-1}(\theta_{k}-\theta_{k+1})H_{k}(p)\right]\\
     &=\theta_{i} \mathbb{E}_{p\sim \mathcal{X}^{*}\left(\theta_{i}\right)}\left[c(p)\right]+\sum_{k=1}^{i-1}(\theta_{k}-\theta_{k+1})\mathbb{E}_{p\sim \mathcal{X}^{*}\left(\theta_{i}\right)}\left[H_{k}(p)\right]\\
     &=\theta_{i} \mathbb{E}_{p\sim \mathcal{X}^{*}\left(\theta_{i}\right)}\left[c(p)\right]+\sum_{k=1}^{i-1}(\theta_{k}-\theta_{k+1})\mathbb{E}_{p\sim \mathcal{X}^{*}\left(\theta_{k}\right)}\left[H_{k}(p)\right]\\
     &=\theta_{i} \mathbb{E}_{p\sim \mathcal{X}^{*}\left(\theta_{i}\right)}\left[c(p)\right]+\sum_{k=1}^{i-1}(\theta_{k}-\theta_{k+1})\mathbb{E}_{p\sim \mathcal{X}^{*}\left(\theta_{k}\right)}\left[c(p)\right]\\
     &=T^{*}(\theta_{i})
 \end{align*} 

 \end{itemize}

 ($\implies$) Suppose that a results-based contract $(\mathcal{X}^{*},t)$ is incentive-compatible and $t$ induces $T^{*}$. We need to show that $\mathcal{X}^{*}$ must be strongly $c$-monotone.

 We claim that such results-based contract must satisfy certain conditions. Next, using the claim, we rewrite the secant hyperplane condition (\ref{eq:shc}).
 
 \begin{claim}\label{claim}
 \begin{align}\label{eq:dif}
     s_{k+1,t}(p)-s_{k,t}(p)=(\theta_{k}-\theta_{k+1})H_{k}(p), \forall p\in\Delta(\Omega) \text{ and }\forall k\in\{1,2,\ldots,N-1\}
 \end{align}
 \end{claim} 

 Firstly, let us consider the type-$\theta_{j}$ researcher does not have incentive to choose posteriors from $S_{i}$:
\begin{align*}
    s_{j,t}(p) \geq u_{j,t}(p):=t(p)-\theta_{j} c(p), \forall p\in S_{i}
\end{align*}

By definition, $s_{i,t}(p)=t(p)-\theta_{i}c(p)$, $\forall p\in S_{i}$. We can substitute $t(p)=s_{i,t}(p)+\theta_{i}c(p)$ into the above inequality and get $s_{j,t}(p) - s_{i,t}(p) \geq \left(\theta_{i}-\theta_{j}\right) c(p)$, $\forall p\in S_{i}$. Then we express $s_{j,t}(p) - s_{i,t}(p)=\sum_{k=i}^{j-1}\left[s_{k+1,t}(p) - s_{k,t}(p)\right]$ as the sum of differences between adjacent types and use (\ref{eq:dif}) to rewrite each term. Because $\theta_{i}-\theta_{j}>0$, dividing it on both sides does not change the sign of the inequality.
\begin{align*}
    \sum_{k=i}^{j-1}\frac{\theta_{k}-\theta_{k+1}}{\theta_{i}-\theta_{j}}H_{k}(p)\geq c(p), \forall p\in S_{i}
\end{align*}

Secondly, consider the type-$\theta_{i}$ researcher does not have incentive to choose posteriors from $S_{j}$:
\begin{align*}
    s_{i,t}(p)\geq t(p)-\theta_{i}c(p), \forall p\in S_{j}
\end{align*}

We can plug in $t(p)=s_{j,t}(p)+\theta_{j} c(p)$ and write it as $\left(\theta_{i}-\theta_{j}\right) c(p) \geq s_{j,t}(p)-s_{i,t}(p)$, $\forall p\in S_{j}$. Similarly, we get   
\begin{align*}
    c(p) \geq \sum_{k=i}^{j-1}\frac{\theta_{k}-\theta_{k+1}}{\theta_{i}-\theta_{j}}H_{k}(p), \forall p\in S_{j}
\end{align*}
 \end{proof}

 \begin{proof}[Proof of Claim \ref{claim}]\hfill

\textbf{Step 1.} We want to show that achieving the lowest payment implies \begin{align*}
    \mathbb{E}_{\mathcal{X}^{*}\left(\theta_{k}\right)}[s_{k+1,t}(p)-s_{k,t}(p)]=\left(\theta_{k}-\theta_{k+1}\right)\mathbb{E}_{\mathcal{X}^{*}\left(\theta_{k}\right)}[ c(p)].
\end{align*}

By assumption that $t$ attains the payment rule $T^{*}$, we have $\mathbb{E}_{\mathcal{X}^{*}(\theta)}[t(p)]=T^{*}(\theta)$, $\forall \theta\in\Theta$.  Because $T^{*}$ is given by (\ref{eq:Tstar1}) and (\ref{eq:Tstar2}), we have for $k \in \{1,2,\ldots,N-1\}$,
\begin{align*}
    \mathbb{E}_{\mathcal{X}^{*}\left(\theta_{k+1}\right)}[t(p)]-\theta_{k+1} \mathbb{E}_{\mathcal{X}^{*}\left(\theta_{k+1}\right)}[c(p)]=\mathbb{E}_{\mathcal{X}^{*}\left(\theta_{k}\right)}[t(p)]-\theta_{k+1} \mathbb{E}_{\mathcal{X}^{*}\left(\theta_{k}\right)}[c(p)]
\end{align*}

By the way $s_{k,t}$ is defined, $s_{k,t}(p)=t(p)-\theta_{k}c(p)$ at $p \in S_{k}$, $\forall k\in \{1,2,\ldots,N\}$. Therefore, we can write $\mathbb{E}_{\mathcal{X}^{*}\left(\theta_{k+1}\right)}[t(p)]=\mathbb{E}_{\mathcal{X}^{*}\left(\theta_{k+1}\right)}[s_{k+1,t}(p)]+\theta_{k+1}\mathbb{E}_{\mathcal{X}^{*}\left(\theta_{k+1}\right)}[c(p)]$ and $\mathbb{E}_{\mathcal{X}^{*}\left(\theta_{k}\right)}[t(p)]=\mathbb{E}_{\mathcal{X}^{*}\left(\theta_{k}\right)}[s_{k,t}(p)]+\theta_{k}\mathbb{E}_{\mathcal{X}^{*}\left(\theta_{k}\right)}[c(p)]$. Plugging these back and simplifying, we get
\begin{align*}
    \mathbb{E}_{\mathcal{X}^{*}\left(\theta_{k+1}\right)}[s_{k+1,t}(p)]=\mathbb{E}_{\mathcal{X}^{*}\left(\theta_{k}\right)}[s_{k,t}(p)]+\theta_{k}\mathbb{E}_{\mathcal{X}^{*}\left(\theta_{k}\right)}[c(p)]-\theta_{k+1} \mathbb{E}_{\mathcal{X}^{*}\left(\theta_{k}\right)}[c(p)]
\end{align*}

By $s_{k+1,t}$ being affine and $\mathbb{E}_{\mathcal{X}^{*}\left(\theta_{k+1}\right)}[p]=\mathbb{E}_{\mathcal{X}^{*}\left(\theta_{k}\right)}[p]=p_{0}$, $\mathbb{E}_{\mathcal{X}^{*}\left(\theta_{k+1}\right)}[s_{k+1,t}(p)]=\mathbb{E}_{\mathcal{X}^{*}\left(\theta_{k}\right)}[s_{k+1,t}(p)]$. Therefore, we can write
\begin{align*}
    \mathbb{E}_{\mathcal{X}^{*}\left(\theta_{k}\right)}[s_{k+1,t}(p)-s_{k,t}(p)]=\mathbb{E}_{\mathcal{X}^{*}\left(\theta_{k}\right)}[\left(\theta_{k}-\theta_{k+1}\right) c(p)]
\end{align*}

\textbf{Step 2.} Next, we will show that the following condition holds: 
\begin{align*}
    s_{k+1,t}(p)-s_{k,t}(p)=\left(\theta_{k}-\theta_{k+1}\right) c(p), \forall p\in S_{k}
\end{align*}

 We prove this by contradiction. Suppose that it is not true. There exists some $q\in S_{k}$ such that $s_{k+1,t}(q)-s_{k,t}(q)<\left(\theta_{k}-\theta_{k+1}\right)c(q)$. Because $s_{k,t}(p)=t(p)-\theta_{k}c(p)$, $\forall p\in S_{k}$, we can write the inequality 
$
    s_{k+1,t}(q)-\left[t(q)-\theta_{k}c(q)\right]<\left(\theta_{k}-\theta_{k+1}\right)c(q)$ equivalently as $ s_{k+1,t}(q)<u_{k+1,t}(q)
$, which contradicts incentive compatibility for the type-$\theta_{k+1}$ researcher in (\ref{eq:shc}).

\textbf{Step 3.} Following Step 2, the affine function $s_{k+1,t}(p)$ is also determined by $|\Omega|$ points in $\{\left(p,u_{k+1,t}(p)\right): p\in S_{k}\}$, so $s_{k+1,t}(p)=\sum_{i=1}^{|\Omega|} \alpha_{i}^{k}\left[u_{k+1,t}(p_{i}^{k})\right]$. By definition, $s_{k,t}(p)=\sum_{i=1}^{|\Omega|} \alpha_{i}^{k}\left[u_{k,t}(p_{i}^{k})\right]$. Moreover, $u_{k+1,t}(p)-u_{k,t}(p)=\left[t(p)-\theta_{k+1}c(p)\right]-\left[t(p)-\theta_{k}c(p)\right]=\left(\theta_{k}-\theta_{k+1}\right)c(p)$. We can rewrite the difference as
\begin{align*}
    s_{k+1,t}(p)-s_{k,t}(p)=\left(\theta_{k}-\theta_{k+1}\right)\sum_{i=1}^{|\Omega|} \alpha_{i}^{k}\left[c(p_{i}^{k})\right]=\left(\theta_{k}-\theta_{k+1}\right) H_{k}(p), 
\end{align*}
which holds for all $p\in \Delta(\Omega)$.

\end{proof}

\begin{proof}[Proof of Proposition \ref{proposition:3}]\hfill

\textbf{Statement 1. A sufficient condition for strong c-monotonicity} 

Consider $i, j\in\{1,2,\ldots,N\}$ and $i<j$. Because $\theta_{1}>\theta_{2}>\ldots>\theta_{N}>0$, for $k\in \{i,i+1,\ldots,j-1\}$, $\frac{\theta_{k}-\theta_{k+1}}{\theta_{i}-\theta_{j}}$ is positive and satisfies $\sum_{k=i}^{j-1}\frac{\theta_{k}-\theta_{k+1}}{\theta_{i}-\theta_{j}}=1$.
    
    Fix $p\in S_{i}$. By definition of $H_{i}$, we have $c(p)= H_{i}(p)$. By condition (\ref{eq:n}), for $k\in \{i+1,i+2,\ldots,j-1\}$, we have $c(p)\leq H_{k}(p)$. Notice that multiplying $\frac{\theta_{k}-\theta_{k+1}}{\theta_{i}-\theta_{j}}$ on both sides will not change the direction of the inequality. Summing up over $k$, we obtain the inequality $c(p)\leq \sum_{k=i}^{j-1}\frac{\theta_{k}-\theta_{k+1}}{\theta_{i}-\theta_{j}}H_{k}(p)$. 
    
    Similarly, fix $p\in S_{j}$. By condition (\ref{eq:n}), for $k\in \{i+1,i+2,\ldots,j-1\}$, we have $c(p)\geq H_{k}(p)$. Summing up over $k$, we obtain $c(p)\geq \sum_{k=i}^{j-1}\frac{\theta_{k}-\theta_{k+1}}{\theta_{i}-\theta_{j}}H_{k}(p)$.

\textbf{Statement 2. A necessary condition for strong c-monotonicity} 

Fix $k\in\{1,2,\ldots, N-1\}$. Setting $i=k$ and $j=k+1$ in the definition of strong $c$-monotonicity, we have $c(p)\geq H_{k}(p)$, $\forall p\in S_{k+1}$.
     Next, we will show that $c(p)\leq H_{k}(p)$, $\forall p\in S_{k-1}$. 
    \begin{itemize}
        \item     If $k>1$, we let $i=k-1$ and $j=k+1$. Strong $c$-monotonicity yields 
    \begin{align*}
        c(p)\leq \frac{\theta_{k-1}-\theta_{k}}{\theta_{k-1}-\theta_{k+1}}H_{k-1}(p)+\frac{\theta_{k}-\theta_{k+1}}{\theta_{k-1}-\theta_{k+1}}H_{k}(p), \forall p\in S_{k-1}
    \end{align*}
    
    By definition, $H_{k-1}(p)=c(p)$, $\forall p\in S_{k-1}$. Replacing $H_{k-1}(p)$ with $c(p)$ on the right-hand side and simplifying, we get $c(p)\leq H_{k}(p)$, $\forall p\in S_{k-1}$. 
    
    \item If $k=1$, $c(p)\leq H_{k}(p)$, $\forall p\in S_{k-1}$ holds vacuously, because we put $S_{0}=\emptyset$.
    \end{itemize}
\end{proof}

\begin{proof}[Proof of Proposition \ref{proposition:2}]\hfill

We will restrict our attention to experiment choice functions that have full dimension for our comparison, as strong $c$-monotonicity is only defined for this class of functions.

\textbf{Statement 1.}
    
Firstly, we want to show that strong $c$-monotonicity implies $c$-monotonicity. 
    
    By Proposition \ref{proposition:3}, Strong $c$-monotonicity implies condition (\ref{eq:nadj}). It follows that for all $k\in \{1,2,\ldots,N-1\}$, $c(p)\geq H_{k}(p)$, $\forall p\in S_{k+1}$. For every $k$, we have
    \begin{align*}
        \mathbb{E}_{p\sim \mathcal{X}^{*}(\theta_{k+1})}[c(p)]\geq& \mathbb{E}_{p\sim\mathcal{X}^{*}(\theta_{k+1})}[H_{k}(p)]\\
        =&H_{k}(\mathbb{E}_{p\sim\mathcal{X}^{*}(\theta_{k+1})}[p])\\
        =&H_{k}(\mathbb{E}_{p\sim\mathcal{X}^{*}(\theta_{k})}[p])\\
        =&\mathbb{E}_{p\sim\mathcal{X}^{*}(\theta_{k})}[H_{k}(p)]\\
        =&\mathbb{E}_{p\sim \mathcal{X}^{*}(\theta_{k})}[c(p)],
    \end{align*}
    \noindent where the first and third equalities follow from $H_{k}$ being affine, the second equality holds because $\mathbb{E}_{p\sim\mathcal{X}^{*}(\theta_{k+1})}[p]=\mathbb{E}_{p\sim\mathcal{X}^{*}(\theta_{k})}[p]=p_{0}$ and the last equality is true by definition $H_{k}(p)=c(p)$, $\forall p \in S_{k}$.
    
    Recall that we write $\mathbb{E}_{p\sim \mathcal{X}^{*}(\theta_{k})}[c(p)]$ succinctly as $C(\mathcal{X}^{*}(\theta_{k}))$. Combining all inequalities, we get $C(\mathcal{X}^{*}(\theta_{N}))\geq C(\mathcal{X}^{*}(\theta_{N-1}))\cdots\geq C(\mathcal{X}^{*}(\theta_{1}))$, meaning that $\mathcal{X}^{*}$ is $c$-monotone.
    
    Next, we want to show that $c$-monotonicity does not imply strong $c$-monotonicity. An example will suffice. 
    
    In the following, we discuss the cases of $|\Omega|=2$ and $|\Omega|>2$.
    
    $\bullet$ For $|\Omega|=2$, we claim that strong $c$-monotonicity is equivalent to Blackwell monotonicity (See the proof of statement 2 below). Any experiment choice function that is $c$-monotone but involves Blackwell incomparable experiments would work.
    
   $\bullet$ For $|\Omega|>2$, we provide a specific example where the experiment choice function that is $c$-monotone but not strong $c$-monotone.

   \begin{example}\label{example:2}

   Define six posterior beliefs as follows.  
\begin{align*}
q_{1}=(0.5770,0.0000,0.4230)\\
q_{1}^{\prime}=(0.0001,0.9998,0.0001)\\
q_{1}^{\prime\prime}=(0.0002,0.0008,0.9990)\\
q_{2}=(0.6799, 0.0001, 0.3200)\\
q_{2}^{\prime}=(0.0005, 0.9993, 0.0002)\\
q_{2}^{\prime\prime}=(0.0004, 0.0170, 0.9827)
\end{align*}
    
$\mathcal{X}(\theta_{1})$ is a distribution over posteriors such that $q_{1}$ occurs with probability $0.58$, $q_{1}^{\prime}$ occurs with probability $0.33$ and $q_{1}^{\prime\prime}$ occurs with probability $0.09$. $\mathcal{X}(\theta_{2})$ is a distribution over posteriors such that $q_{2}$ occurs with probability $0.49$, $q_{2}^{\prime}$ occurs with probability $0.33$ and $q_{2}^{\prime\prime}$ occurs with probability $0.18$. $\mathcal{X}$ has full dimension.

Consider a quadratic cost function $$c(p)=\left(p\left(\omega_{1}\right)-\frac{1}{3}\right)^{2}+\left(p\left(\omega_{2}\right)-\frac{1}{3}\right)^{2}+\left(p\left(\omega_{3}\right)-\frac{1}{3}\right)^{2}.$$ 

Firstly, $\mathcal{X}$ is $c$-monotone, because $C(\mathcal{X}(\theta_{1}))=0.3832<0.44678=C(\mathcal{X}(\theta_{2}))$. Secondly, $\mathcal{X}$ is not strongly $c$-monotone. With two types, strong $c$-monotonicity is equivalent to $c(p)\geq H_{1}(p)$, $\forall p\in \support(\mathcal{X}(\theta_{2}))$. We find $q_{2}^{\prime\prime}\in \support(\mathcal{X}(\theta_{2}))$ but $c(q_{2}^{\prime\prime})\geq H_{1}(q_{2}^{\prime\prime})$, so $\mathcal{X}$ is not strongly $c$-monotone.  

  \end{example}

    As preparation for proving statements 2 and 3, we state a geometric version of Blackwell's theorem in \cite{W2018} that holds for any finite state space. 

        \begin{lemma}\label{lemma:3} (\cite{W2018})
        For $\tau$, $\tau^{\prime}\in X$ and $\tau^{\prime}$ has affinely independent support, $\tau^{\prime}$ is a mean-preserving spread of $\tau$ if and only if $\support(\tau)\subseteq \convexhull(\support(\tau^{\prime}))$.
    \end{lemma}
    
    \textbf{Statement 2.} 

    For $|\Omega|=2$: We will first show $D_{k}=\convexhull(S_{k})$ and then use Proposition \ref{proposition:3} and Lemma \ref{lemma:3} to connect strong $c$-monotonicity with Blackwell monotonicity.  
    
    Note that by strict convexity of $c$, $\convexhull(S_{k})\subseteq D_{k}$. We will prove $D_{k} \subseteq \convexhull(S_{k})$ by contradiction. Suppose there is $q\in \Delta(\Omega)$ such that $q\in D_{k}$ and $q\not\in \convexhull(S_{k})$. Specifically, write $S_{k}=\{p_{k},p_{k}^{\prime}\}$. By affine independence of $S_{k}$, there exists $\lambda \in \mathbb{R}$ such that $q=\lambda p_{k}+(1-\lambda)p_{k}^{\prime}$. Because $q\not\in \convexhull(S_{k})$, $\lambda\not\in[0,1]$.

    \begin{itemize}
        \item If $\lambda<0$, then $1-\lambda>0$. 
        \begin{align*}
            &q=\lambda p_{k}+(1-\lambda)p_{k}^{\prime}\\
            \iff \quad &(1-\lambda)p_{k}^{\prime}=q-\lambda p_{k}\\
            \iff \quad &p_{k}^{\prime}=\frac{1}{1-\lambda} q + \frac{-\lambda}{1-\lambda} p_{k}\\
            \iff \quad &p_{k}^{\prime} \in \convexhull(\{q,p_{k}\})
        \end{align*}
        We reach a contradiction, since
    \begin{align*}
        c(p_{k}^{\prime})<\frac{1}{1-\lambda} c(q)+\frac{-\lambda}{1-\lambda} c(p_{k})\leq \frac{1}{1-\lambda} H_{k}(q)+\frac{-\lambda}{1-\lambda} c(p_{k})\\
        =\frac{1}{1-\lambda} H_{k}(q)+\frac{-\lambda}{1-\lambda} H_{k}(p_{k})=
        H_{k}(p_{k}^{\prime})=c(p_{k}^{\prime}),
    \end{align*}
 \noindent where the strict inequality follows from the strict convexity of $c$, the weak inequality is due to $q\in D_{k}$, the first and third equalities hold by $H_{k}(p)=c(p)$ at every $p\in S_{k}$ and the second equality holds because $H_{k}$ is affine.
        \item If $\lambda>1$, then $1-\lambda<0$. 
        \begin{align*}
            &q=\lambda p_{k}+(1-\lambda)p_{k}^{\prime}\\
            \iff \quad &\lambda p_{k}=q-(1-\lambda)p_{k}^{\prime}\\
            \iff \quad &p_{k}=\frac{1}{\lambda} q + \frac{-(1-\lambda)}{\lambda} p_{k}^{\prime}\\
            \iff \quad &p_{k} \in \convexhull(\{q,p_{k}^{\prime}\})
        \end{align*}
Similarly, we reach a contradiction because
\begin{align*}
        c(p_{k})<\frac{1}{\lambda} c(q) + \frac{-(1-\lambda)}{\lambda} c(p_{k}^{\prime}) \leq \frac{1}{\lambda} H_{k}(q) + \frac{-(1-\lambda)}{\lambda} c(p_{k}^{\prime})=c(p_{k}).
    \end{align*}
        
    \end{itemize}
    
    Next, we will verify that (i) Blackwell monotonicity is implied by the necessary condition for strong $c$-monotonicity and (ii) Blackwell monotonicity implies the sufficient condition for strong $c$-monotonicity. Thus, all of these are equivalent.
    
    (i) By condition (\ref{eq:nadj}) and $D_{k}=\convexhull(S_{k})$, $S_{k-1}\subseteq \convexhull(S_{k})$, $\forall k\in\{1,2,\ldots, N-1\}$. It implies Blackwell monotonicity, because the Blackwell order is transitive.
    
    (ii) By Blackwell monotonicity, $S_{j}\subseteq \convexhull(S_{i})$, $\forall i,j\in \{1,2,\ldots, N\}$ and $j<i$. Because $D_{k}=\convexhull(S_{k})$, we immediately have $S_{i}\subseteq D_{k}$, $\forall i<k$. 
    
    The rest of the proof is to show $S_{i}\subseteq U_{k}$, $\forall i>k$. We will prove this by contradiction. Write $S_{i}=\{p_{i},p_{i}^{\prime}\}$. Suppose that $p_{i}\not\in U_{k}$. It follows that $p_{i}\in D_{k}=\convexhull(S_{k})$ and $p_{i}\not \in S_{k}$. Thus, there is $\lambda\in(0,1)$ such that $p_{i}=\lambda p_{k}+(1-\lambda) p_{k}^{\prime}$. By affine independence of $S_{k}$, there is $\mu\in \mathbb{R}$ and $\mu\neq \lambda$ such that $p_{i}^{\prime}=\mu p_{k}+(1-\mu) p_{k}^{\prime}$. We can write $p_{k}$ and $p_{k}^{\prime}$ as affine combinations of $p_{i}$ and $p_{i}^{\prime}$:
\begin{align*}
    \begin{pmatrix}
p_{i} \\
p_{i}^{\prime}
\end{pmatrix} =
    \begin{pmatrix}
\lambda & 1-\lambda \\
\mu & 1-\mu
\end{pmatrix} 
    \begin{pmatrix}
p_{k} \\
p_{k}^{\prime}
\end{pmatrix} \implies     \begin{pmatrix}
p_{k} \\
p_{k}^{\prime}
\end{pmatrix} =
    \begin{pmatrix}
\frac{1-\mu}{\lambda-\mu} & \frac{\lambda-1}{\lambda-\mu} \\
\frac{-\mu}{\lambda-\mu} & \frac{\lambda}{\lambda-\mu}
\end{pmatrix} 
    \begin{pmatrix}
p_{i} \\
p_{i}^{\prime}
\end{pmatrix} 
\end{align*}

    By affine independence of $S_{i}$, the representation of an affine combination is unique. Because $\lambda(\lambda-1)<0$, we can never have both $p_{k}$ and $p_{k}^{\prime}$ belong to $\convexhull(S_{i})$. However, this contradicts the fact that $S_{k}\subseteq \convexhull(S_{i})$, for $i>k$.
   
    \textbf{Statement 3.}
    
    For $|\Omega|>2$: Example \ref{example:1} shows that for a fixed $c$, strong $c$-monotonicity does not guarantee Blackwell monotononicity. Towards the other direction, it is possible to find a Blackwell-monotone choice function that violates strong $c$-monotonicity with some properly chosen function $c$. (The idea is to draw a convex set $D_{1}$ that includes one of the posteriors in $S_{2}$.)
\end{proof}

\section{Relation to a general result in Yoder (2022)} \label{section:appB}
 
Proposition 4 in \cite{Y2022} contains a result that holds for any finite state space. Firstly, it provides a necessary condition for incentive compatibility in (\ref{newic2}). Secondly, it can be applied to strengthen our Proposition \ref{proposition:1}.       
 
Recall that the optimal experiment choice function can be obtained from solving a set of type-specific problems (\ref{eq:type-specific}). The objective associated with a more efficient type is more convex than the one associated with a less efficient type, in the sense that their difference is strictly convex. It follows from \cite{Y2022} that 
\begin{align}\label{eq:comp}
    \forall i<j:~ S_{j} \cap \operatorname{conv}\left(S_{i}\right) \subseteq \operatorname{ext}\left(\operatorname{conv}\left(S_{i}\right)\right)
\end{align}
where $\operatorname{ext}\left(\operatorname{conv}\left(S_{i}\right)\right)$ represents the set of extreme points of $\operatorname{conv}\left(S_{i}\right)$. To interpret, all of the results produced by a more efficient researcher (with type $\theta_{j}$) must be at least as extreme as any result produced by a less efficient researcher (with type $\theta_{i}$). Also, it is known that this ordering is equivalent to the Blackwell ordering when $|\Omega|=2$, but in general is neither stronger nor weaker. 
 
Under the full dimension assumption, we have a result showing that strong $c$-monotonicity is strictly stronger than condition (\ref{eq:comp}). 
 
\begin{proposition}
Suppose that $|\Theta|=2$. Strong $c$-monotonicity implies (\ref{eq:comp}); while the converse is not true.
\end{proposition}
 
\begin{proof}
Assuming that strong $c$-monotonicity holds, i.e.,  $S_{2}\subseteq U_{1}$, we have $$\left(S_{2}\cap \convexhull (S_{1})\right)\subseteq \left(U_{1}\cap \convexhull (S_{1})\right)=S_{1}=\operatorname{ext}\left(\operatorname{conv}\left(S_{1}\right)\right).$$ 

Conversely, Example \ref{example:2} shows that given the quadratic function $c$, the experiment choice function is not strongly $c$-monotone. However, it satisfies condition (\ref{eq:comp}), because $S_{2}\cap \convexhull (S_{1})=\emptyset$.
\end{proof}

To emphasize, strong $c$-monotonicity depends on the cost function $c$ while the ordering of experiments in (\ref{eq:comp}) does not. Typically, if an experiment choice function satisfies condition (\ref{eq:comp}), it is not necessary to have the outcome equivalence between results-based contracting and methods-based contracting.

\end{document}